\documentclass[article, shortnames, nojss]{jss}
\usepackage{hyperref}
\usepackage{amsmath,amssymb,amsfonts}
\usepackage{microtype}
\usepackage{algorithm}
\usepackage{algpseudocode}
\usepackage{array}
\usepackage{comment}
\usepackage{makecell}
\usepackage{natbib}
\usepackage{bbding}
\usepackage{bm}
\usepackage[capitalise]{cleveref}

\def\r{{\mathbf r}}
\def\q{{\mathbf q}}
\def\v{{\mathbf v}}
\def\x{{\mathbf x}}
\def\y{{\mathbf y}}
\def\z{{\mathbf z}}

\def\v{{\mathbf v}}
\def\f{{\mathbf f}}
\def\g{{\mathbf g}}
\def\w{{\mathbf w}}

\def\A{{\mathbf A}}
\def\H{{\mathbf H}}
\def\X{{\mathbf X}}
\def\Y{{\mathbf Y}}
\def\R{{\mathbf R}}
\def\Q{{\mathbf Q}}
\def\C{{\mathbf C}}

\newcommand{\reals}{\mathbb R}      

\newcommand{\Input}{\State\textbf{Input: }}
\newcommand{\Output}{\State\textbf{Output: }}

\newcommand{\bra}[1]{\left({#1}\right)}
\newcommand{\sbra}[1]{\left[{#1}\right]}
\newcommand{\cbra}[1]{\left\{{#1}\right\}}

\newcommand{\upindex}[1]{^{\bra{#1}}}

\newcommand{\grad}{\nabla_{\bm\theta}}
\newcommand{\expec}{\mathbb E}
\newcommand{\norm}[1]{\left\lVert#1\right\rVert}

\newcolumntype{?}{!{\vrule width 1pt}}
 
\author{John-Joseph Brady\\King's College \\London \And Jessica Cox\\Max-Planck\\Institut f\"ur Physik \And Yunpeng Li \\ King's College \\ London\\ \And V\'ictor Elvira \\University of \\Edinburgh}
\title{\pkg{PyDPF}: A \proglang{Python} Package for Differentiable Particle Filtering}

\Plainauthor{John-Joseph Brady, Jessica Cox, V\'ictor Elvira, Yunpeng Li} 
\Plaintitle{PyDPF: a python package for differentiable particle filtering} 
\Shorttitle{\pkg{PyDPF}} 

\Abstract{
    State-space models (SSMs) are a widely used tool in time series analysis. 
    In the complex systems that arise from real-world data, it is common to employ particle filtering (PF), an efficient Monte Carlo method for estimating the hidden state corresponding to a sequence of observations.
    Applying particle filtering requires specifying both the parametric form and the parameters of the system, which are often unknown and must be estimated.
    Gradient-based optimisation techniques cannot be applied directly to standard particle filters, as the filters themselves are not differentiable.
    However, several recently proposed methods modify the resampling step to make particle filtering differentiable.
    In this paper, we present an implementation of several such differentiable particle filters (DPFs) with a unified API built on the popular \pkg{PyTorch} framework.
    Our implementation makes these algorithms easily accessible to a broader research community and facilitates straightforward comparison between them.
    We validate our framework by reproducing experiments from several existing studies and demonstrate how DPFs can be applied to address several common challenges with state space modelling.
}
\Plainkeywords{differentiable particle filter, state-space model, Python} 

\Address{
    John-Joseph Brady\\
    Centre for Oral, Clinical \& Translational Studies\\
    Faculty of Dentistry, Oral \& Craniofacial Sciences\\
    King's College London\\
    London, United Kingdom\\
    E-mail: \email{john-joseph.brady@kcl.ac.uk}\\\vspace{1em}\\
    Jessica Cox\\
    Max-Planck-Institut f\"ur Physik\\
    Garching, Germany\\
    E-mail: \email{bcox@mpp.mpg.de}\\
    ORCID: \href{https://orcid.org/0000-0002-0718-4450}{0000-0002-0718-4450}\\\vspace{1em}\\
    Yunpeng Li\\
    Centre for Oral, Clinical \& Translational Studies\\
    Faculty of Dentistry, Oral \& Craniofacial Sciences\\
    King's College London\\
    London, United Kingdom\\
    E-mail: \email{yunpeng.li@kcl.ac.uk}\\\vspace{1em}\\
    V\'ictor Elvira\\
    School of Mathematics\\
    University of Edinburgh\\
    Edinburgh, United Kingdom\\
    E-mail: \email{victor.elvira@ed.ac.uk}\\
}

\begin{document}

\section{Introduction}
    State-space models are a powerful statistical framework for analysing sequential data.
    In these models, the system is modelled via a sequence of unobserved latent states that evolve in time, which are related to a sequence of noisy observations.
    These models have been used in many fields, such as target tracking \citep{wang2017survey}, finance \citep{virbickaite2019particle}, epidemiology \citep{chen2011tracking}, ecology \citep{newman2023state}, and meteorology \citep{clayton2013operational}.
    Given a state-space model, it is commonly required to estimate the underlying hidden state conditional on the sequence of observations obtained until a given time, a task known as the filtering problem.
    If the state and observation models are linear with Gaussian noise, a model known as the \emph{linear-Gaussian state-space model}, the filtering problem can be optimally solved via the Kalman filter \citep{kalman1960new}.
    However, if the dynamics are non-linear, or the noise distribution is non-Gaussian, we must use approximate filters.
    Methods such as the extended Kalman filter and the unscented Kalman filter approximate the filtering posterior with a Gaussian, which can lead to inaccurate results.
    An alternative is sequential Monte Carlo (SMC), also known as particle filtering, which approximates the filtering distributions using a set of Monte Carlo samples.
    These methods are the focus of this work.
    
    All filtering methods require that the parameters of the state-space model are either known or suitably estimated.
    Since, in general, the parameters of the model are not known, we must estimate them using the information contained in the observation series.
    In the case of the Kalman filter, the parameter likelihood can be obtained exactly, and an expectation-maximisation scheme can be applied to jointly estimate the unknown parameters of a linear-Gaussian state-space model \citep{sarkka2023bayesian}.
    However, when a particle filter is applied, the parameter likelihood can only be estimated.
    Furthermore, particle filters do not admit gradients of the parameter likelihood with respect to the parameters, as the parameter likelihood is estimated via the particle weights. These weights are the result of non-differentiable operations, namely their dependence on the parameters is, in part, through the sampling of a categorical distribution.
    To address this limitation, several methods have recently been designed, collectively termed \emph{differentiable particle filters} (DPFs), which aim to make the particle filter differentiable with respect to its input parameters.

    In this paper, we present our unified implementation of several DPFs, including \citep{jonschkowski2018DPF, karkus2018DPF, Corenflos2021OT, scibior2021stopgrad, younis2024mixtureresample}, in the \proglang{Python} package \pkg{PyDPF} (\textbf{Py}thon \textbf{D}ifferentiable \textbf{P}article \textbf{F}iltering).
    The implementation is designed to be simple to use, extensible, and efficient, and is aimed at both the particle filter community and the wider scientific community.
    Our framework allows for rapid development of modified particle filters, easy benchmarking of different DPF algorithms, and makes it simple to use state-of-the-art differentiable particle filters on a user-defined problem.
    To the best of our knowledge, this is the first implementation of such a framework.
    Our package and code to run the experiments is available at the \pkg{PyDPF} repository.\footnote{\url{https://github.com/John-JoB/pydpf}} \pkg{PyDPF} may be installed from pypi using the \pkg{pip} package manager with the command \code{pip install pydpf}. Complete documentation for \pkg{PyDPF} can be found on readthedocs.\footnote{\url{https://python-dpf.readthedocs.io/en/latest/}} 
    We provide several example experiments, which demonstrate to the user the entire process of using \pkg{PyDPF}, from loading data to learning parameters.

\subsection{Overview of features}
The core contribution of \pkg{PyDPF} is that it provides a framework for the development and implementation of differentiable particle filters (see \cref{sec:dpfs} or \citet{Chen2025review} for an overview). We achieve this by building a particle filtering package inside an auto-differentiation engine, specifically \pkg{PyTorch} which we endow with a ``plug-and-play" design in both the statistical models under study and the inference methodology itself. We demonstrate this functionality in \cref{sec:dpfs} by implementing various popular DPF algorithms within our framework, and package them with PyDPF so that they also might be used out-of-the-box by practitioners. Developing a common framework for DPFs facilitates rapid testing of algorithms; given a parametric model a modeller can evaluate the performance of a range of DPFs on their specific problem with only a few lines of code for each methodology. We demonstrate this by benchmarking the implemented algorithms on a variety of tasks in \cref{sec:ex}.

Our choice to build our package with \pkg{PyTorch} means that the implementation flow of parameter inference is very familiar to the deep learning community (see \cref{sec:ex_usage}). Furthermore, it makes integration with other packages in the mature \pkg{PyTorch} ecosystem whether as part of statistical model, as part of the filtering methodology, or to process experimental results simple.

\subsection{Comparison to existing software packages}
    To the best of our knowledge, \pkg{PyDPF} is the first published software package, in any language, designed to support the development and deployment of DPFs. No other package implements a broad range of DPFs. Several packages are available for standard particle filtering, such as \pkg{pypfilt}\footnote{\url{https://gitlab.unimelb.edu.au/rgmoss/particle-filter-for-python}} \citep{moss2024pypfilt} in \proglang{Python}, \pkg{LowLevelParticleFilters.jl}\footnote{\url{https://github.com/baggepinnen/LowLevelParticleFilters.jl}} \citep{lowlevelpf} in \proglang{Julia}, and the Control system toolbox\footnote{\url{https://uk.mathworks.com/help/control/ref/particlefilter.html}} \citep{matlabcst} in \proglang{MATLAB}. There are also packages for parameter inference in particle filters using gradient-free methods, such as \pkg{pomp}\footnote{\url{https://github.com/kingaa/pomp}} \citep{king2016pomp, pompmanual} in \proglang{R}, or \pkg{Turing.jl}\footnote{\url{https://github.com/TuringLang/Turing.jl}} \citep{ge2018turing} in \proglang{Julia}.
    Some packages implement specific differentiable particle filters, such as the optimal transport resampling DPF implemented in the \pkg{FilterFlow}\footnote{\url{https://github.com/JTT94/filterflow}} \citep{Corenflos2021OT} \proglang{Python} package. Gradient based training of specific state space models \citep{krishnan2015dkf} is implemented in the \pkg{deep-kalman-filter}\footnote{\url{https://github.com/kokikwbt/deep-kalman-filter}} \proglang{Python} package \citep{kawabata2022dkf}.

     State space models are defined in \pkg{PyDPF} very similarly to all other mentioned state space model specific packages (\pkg{pypfilt}, \pkg{LowLevelParticleFilters.jl}, \proglang{MATLAB}'s Control system toolbox, \pkg{pomp}, and \pkg{FilterFlow}) in that the user specifies a transition rule and an observation dependent weighting function. However, unlike all other mentioned packages, we extend the plug-and-play paradigm to resampling algorithms and particle proposal distributions, thereby enabling ready access to state-of-the-art approaches in classical and differentiable filtering. Furthermore, \pkg{PyDPF} is designed with researchers working on methodological innovation in mind, therefore we have designed our package to serve as a development framework for users to build upon. We demonstrate this functionality with the implementation of several published DPFs that we include in our package. \pkg{PyDPF}'s close adherence to \pkg{PyTorch} conventions makes it easily accessible to the deep learning community and allows simple integration of the myriad deep neural models implemented in \pkg{PyTorch} into a particle filter or DPF.

    Parameter inference in \pkg{PyDPF} is focussed solely on gradient based optimisation. We do not include gradient free optimisation approaches such as those in \pkg{LowLevelParticleFilters.jl} and \pkg{pomp}. Nor does \pkg{PyDPF} provide any functionality for Bayesian parameter estimation approaches such as particle Markov chain Monte Carlo \cite{andrieu2010particle}, which is implemented in \pkg{pomp} and \pkg{Turing.jl}. Furthermore, we do not include any algorithms for the smoothing of state space trajectories.
    

    \subsection{Paper overview}
    The remainder of this paper is structured as follows.
    In Section \ref{sec:pydpf_basics}, we cover the base structure and conventions of our package necessary to understand the code snippets throughout the paper.
    In Section \ref{sec:ssms}, we provide the background information of state-space models and particle filters.
    Section \ref{sec:dpfs} introduces differentiable particle filters and techniques used to make particle filters differentiable.
    We discuss in Section \ref{sec:imp-alg} the methods that we implement in \pkg{PyDPF}, and provide examples of their use.
    Section \ref{sec:ex_usage} consolidates the previous chapters to provide a full walk-through of the process of setting up and running a simple machine learning experiment in \pkg{PyDPF}.
    We then discuss advanced usage of \pkg{PyDPF} in Section \ref{sec:adv_usage}.
    In Section \ref{sec:ex} we present a set of example experiments that simultaneously demonstrate a range of tasks to which \pkg{PyDPF} can be applied; and, by comparison to oracle approaches and the replication of previous studies, validate our implementation.
    We provide concluding remarks in Section \ref{sec:conc}.

\section{PyDPF basics}
\label{sec:pydpf_basics}
    Before we present the functionality provided by our package, we review the basic structures and conventions we employ in \pkg{PyDPF} that a user needs to be familiar with. A key design consideration in \pkg{PyDPF} is extensibility. In this paper, we outline existing algorithms that we have implemented and included in the package, but we also envision researchers extending \pkg{PyDPF} to suit their own needs. The package is designed to integrate as seamlessly as possible with base \pkg{PyTorch} \citep{paszke2019PyTorch}. Following typical \pkg{PyTorch} design patterns, \pkg{PyDPF} is rigidly object-oriented. 
\subsection{PyDPF Modules}
    \pkg{PyDPF} includes its own \code{Module} class that extends \code{torch.nn.Module} that we find useful in defining custom parameterised probability distributions. We include the following two \code{Property}-like environments.\par
    \textbf{cached\_property:} Used to cache the results of functions of the parameters. For example if we need to repeatedly use the matrix inverse of a parameter. Gradients can be passed through the transform that creates the \code{cached\_property}. Cached\_properties can be stacked.

    \textbf{constrained\_parameter:} Used to constrain parameters. \code{constrained\_parameter} applies a transform in-place from an unconstrained parameter to a parameter satisfying the required constraint. Because the operation is in-place, gradient tracking through this transform is not supported. It is intended to prevent parameters from entering disallowed regions, such as ensuring a variance remains positive. There should exist an allowed region where the parameter remains unchanged. Because the modifications are made in-place, \code{constrained\_parameter} objects cannot depend on other \code{constrained\_parameter} objects or \code{cached\_property} objects. If it is desired to constrain functions of the parameters rather than the parameters themselves, we recommend \pkg{PyTorch}'s \code{parametrize} API which provides similar functionality but operates out-of-place.

    We provide the following minimal example that shows how one might implement a \pkg{PyDPF} module that evaluates the probability density function of a Gaussian, where the mean and variance are parameters.

\footnotesize
    \begin{Code}
class GaussianDensity(pydpf.Module):
    
    log_2pi = math.log(2*math.pi) 
    
    def __init__(self, initial_mean, initial_variance):
        super().__init__()
        self.mean = torch.nn.parameter.Parameter(torch.tensor(initial_mean))
        self.variance_data = torch.nn.parameter.Parameter(torch.tensor(initial_variance))
        
    @pydpf.constrained_parameter
    def variance(self):
        #Return references to the parameter we want to modify in place and 
        #to a tensor containing the new value 
        return self.variance_data, torch.abs(self.variance_data)
    
    @pydpf.cached_property
    def inverse_variance(self):
        return 1/self.variance
    
    @pydpf.cached_property
    def log_variance(self):
        return torch.log(self.variance)
        
    def log_density(self, input):
        sqd_residual = (input - self.mean)**2
        return -(sqd_residual*self.inverse_variance + self.log_2pi + self.log_variance)/2
    \end{Code}
\normalsize
    \pkg{PyTorch} does not provide a built-in way to detect optimiser steps, so we have to manually update the model by calling \code{.update()} on the highest level \code{Module} in a model whenever the parameters may have changed. For this reason, if any sub-modules in a model have either a \code{cached_property} or a \code{constrained_parameter} the top-level module should be a \code{pydpf.Module}. However, any sub-module can be a \code{torch.nn.Module} without consequence.

\subsection{PyDPF data categories}

\begin{table}[t]
        \centering
        \begin{tabular}{|c|m{15em}|c|}
            \hline
             Label&\centering Usage&Data type\\
             \hline
             \hline
             \code{state}&The particle estimates of the latent state of the state-space system at the current time-step&Tensor $(B\times K \times D_{s})$\\
             \hline
             \code{weight}&The log weights of the particles, entries aligned to \code{state}&Tensor $(B\times K)$\\
             \hline
             \code{prev\_state}&The particle estimates of the latent state of the state-space system at the previous time-step&Tensor $(B\times K \times D_{s})$\\
             \hline
             \code{observation}&The observations of the state-space system at the current time-step& Tensor $(B\times D_{o})$\\
             \hline
             \code{control}&Control actions or other exogenous variables at the current time-step&Tensor $(B\times D_{c})$\\
             \hline
             \code{time}&The time the current time-step occurs at&Tensor $(B)$\\
             \hline
             \code{prev\_time}&The time of the previous time-step&Tensor $(B)$\\
             \hline
             \code{series\_metadata}&Exogenous variables that are constant for a given trajectory&Tensor $(B\times D_{m})$\\
             \hline
             \code{t}&The index of the time-step&Integer\\
             \hline
        \end{tabular}
        \caption{The intended usage of the included data-categories that can be passed as arguments to user defined functions.}
        \label{tab:data_cats}
    \end{table}

    The intended usage of \pkg{PyDPF} is for users to create their own models and algorithms as \code{pydpf.Module} objects and to define custom functions to interface with those provided by the package. To facilitate this, a schema is needed for the different variables passed from the base filtering algorithms to user-defined functions. Table \ref{tab:data_cats} defines this schema. \pkg{PyDPF} follows a batch-sequential paradigm, with additional dimensions for each sample drawn, known in the SMC literature as a particle, and the intrinsic dimensionality of the distribution. Tensors handled and returned by \pkg{PyDPF} functions have the shape $\bra{T\times B \times K \times D_{(\cdot)}}$ corresponding to (time-step $\times$ batch $\times$ particle $\times$ intrinsic-dimension). Frequently, one or more of these dimensions will not be present, in which case ordering is maintained. For example, the observations are independent of the particle index and vary only with time-step and batch. So, the inputted observation tensor has the shape $\bra{T \times B \times D_{o} }$, where $D_o$ is the dimension of the observations. 
    
    When we pass the data as arguments to user-defined functions, a single time-step is indexed. Therefore, the data-types and dimensions described in Table \ref{tab:data_cats} are what the user-defined functions will receive. In \pkg{PyDPF}, all arguments are passed by keyword, so any unnecessary arguments received from a \pkg{PyDPF} call to a user-defined function can be conveniently grouped into a \code{**dictionary} object.

    The tensor shapes given in Table~\ref{tab:data_cats} are accurate for the tensors as they are passed to any user defined functions. When passing data to the filtering algorithm, aside from \code{series_metadata}, they should have an additional dimension for the time-step.
    \code{prev_state}, \code{prev_time} and \code{t} are calculated automatically so don't need to be passed.

\subsection{PyDPF deserialisation and data loading}
\label{sec:basic:data}

    For convenience, \pkg{PyDPF} provides methods to load data from files into a map-style \\ \code{torch.utils.data.Dataset} object. Data can be stored in one of two formats, either the entire dataset in a single \code{.csv} file, or each trajectory in separate files named \code{\{1.csv, 2.csv, \dots, T.csv\}} in a dedicated directory. These .csv files are formed of headed columns and there must be at least one \code{observation} column, with \code{state}, \code{time}, and \code{control} columns being optional. As all the data categories, apart from \code{time}, are vector valued there can be multiple columns for each category corresponding to each of the $D_{(\cdot)}$ dimensions. For the single-file format there must be additionally a \code{series_id} column that will be used to index each trajectory, for the multiple file format the \code{series_id} is encoded in the file name.

    The data category \code{series_metadata} exists to store exogenous variables that the trajectories might depend on, but are constant over a trajectory. If \code{series_metadata} is required it should be stored in a separate \code{.csv} indexed by a \code{series_id} column.

    Given a file in the required format, loading a dataset is simple: call \code{pydpf.StateSpaceDataset} with the data path, the column prefixes and the device to store data retrieved by the data loader, see the code example below. 

\footnotesize
\begin{Code}
dataset = pydpf.StateSpaceDataset(data_path=data_path, 
                                  series_id_column='series_id', 
                                  state_prefix='state', 
                                  observation_prefix='observation', 
                                  device='cpu')
data_loader = torch.utils.data.DataLoader(dataset, 
                                          batch_size=10, 
                                          shuffle=True,  
                                          collate_fn=dataset.collate)
\end{Code}
\normalsize

When initialising the data loader, it is crucial that the argument \code{collate_fn} is set to \code{dataset.collate} where \code{dataset} is the dataset passed to the data loader. \pkg{PyTorch}'s default collate function will not return the data in a format that obeys \pkg{PyDPF} conventions. When looping over the data loader, data is returned as tuple in the ordering \code{state} - \code{observation} - \code{time} - \code{control} - \code{series_metadata} with only the fields that are present in the dataset being returned.

\subsection{Reproducibility}
\label{sec:basics:reproducibility}
    \pkg{PyTorch} does not provide the fine-grained tracking of pseudo-random state offered by competing numerical libraries such as \pkg{JAX} \citep{jax2018github}. Our approach, used in all built-in implementations with pseudo-random operations, and that we recommend the user adopt for all their extensions, is to initialise a random generator per Module that is used to control all random operations used within that Module.

    Some torch \pkg{CUDA} operations are non-deterministic by default. Refer to the documentation of \code{torch.use_deterministic_algorithms}\footnote{\url{https://docs.pytorch.org/docs/stable/generated/torch.use_deterministic_algorithms.html}} for detail. This non-determinacy is at the order of precision over a single operation, but our tests showed it can result in a significant variance over the course of a full forward pass. To mitigate this, we provide a context manager \code{pydpf.utils.set_deterministic_mode(mode, warn_only)} that, when \code{mode = True} sets the environment variable
    
    \code{"CUBLAS_WORKSPACE_CONFIG" = ":4096:8"} 
    
    and calls 
    
    \code{torch.use_deterministic_algorithms(True, warn_only=warn_only)} 
    
    before reverting to default settings on context exit. Under this context we expect an increase in the time and memory costs for \pkg{CUDA} operations compared to the non-deterministic implementations.
    
    Note, however, that several of the implemented DPFs rely on the \code{torch.cumsum()} operation that is not guaranteed to be deterministic even under the \small \code{pydpf.utils.set_deterministic_mode} \normalsize context manager. Despite this, in our experiments we observe that the results are consistent on our set-up. Furthermore, \pkg{PyTorch} does not guarantee reproducibility across different hardware, \pkg{PyTorch} versions or versions of upstream dependencies such as \pkg{CUDA}. For this reason we repeat all our experiments across several random seeds to mitigate some of this unavoidable variance.

\section{Background}
    \label{sec:ssms}
\subsection{State-space models}
    \label{sec:ssms:ssms}
    State-space models (SSMs) are used to model temporally varying systems via a hidden state.
    A general state-space model is given by
    \begin{equation}
        \begin{alignedat}{2}
            \x_t &\sim p(\x_t | \x_{t-1}; \bm\theta),\\
            \y_t &\sim p(\y_t | \x_t; \bm\theta),
        \end{alignedat} \label{eq:ssms:ssm}
    \end{equation}
    where $t \in \{1,\dots, T\}$ denotes discrete time, $\x_t \in \reals^{d_x}$ is the hidden state of the system at time $t$, $\y_t \in \reals^{d_y}$ is the observation associated with $\x_t$, $\bm\theta$ is a set of parameters relating to the system dynamics, and the Markov kernels $p(\x_t | \x_{t-1}; \bm\theta)$ and $p(\y_t | \x_t; \bm\theta)$ encode the transition and observation model respectively.
    The initial value of the state, $\x_0$, is distributed $\x_0 \sim p(\x_0|\bm\theta)$.
    Note that state-space models are Markov in the state, meaning that $p(\x_t|\x_{0:t-1}; \bm\theta) = p(\x_t|\x_{t-1}; \bm\theta)$, where here and throughout this paper we use the index slice notation $a_{\alpha:\beta} = \cbra{\alpha_{i}}^{\beta}_{i=\alpha}$. Furthermore, the observation at $t$ depends only on the state at $t$, meaning that $p(\y_t|\x_{0:t}) = p(\y_t|\x_t)$. Formally all of these distributions may depend on $t$ and any other a priori known set of constants. In \pkg{PyDPF} SSMs may depend on such arbitrary constants through the data categories of \code{control}, \code{time}, \code{prev_time}, \code{t}, and \code{series_metadata}, see Table \ref{tab:data_cats}. For clarity, we keep this dependence implicit in our notation.

    The sequence of hidden states, $\x_{0:T}$, is typically unobserved. Instead, we observe the sequence of related measurements $\y_{0:T}$.
    It is commonly required to infer the hidden states conditional on the observation sequence.
    When this inference is performed such that $\x_t$ is inferred using only $\y_{0:t}$, it is called the filtering problem, and $p(\x_t | \y_{0:t})$ is known as the filtering distribution.
    In certain specific cases of state-space models, there exist closed-form solutions to the filtering problem, such as the Kalman filter \citep{kalman1960new}.
    However, for most state-space models, there are no closed-form solutions to the filtering problem, and we must rely on approximate inference methods.

\subsubsection{Defining a model in PyDPF}
\label{sec:ssms:ssms:imp}
In \pkg{PyDPF} state-space models are most naturally defined as \pkg{PyDPF} (or \pkg{PyTorch}) Modules. The methods required by each Module depend on its intended use. In this section, we describe the functions needed to support all filtering algorithms in \pkg{PyDPF}. Tables \ref{tab:prior_methds}-\ref{tab:prop_methds} summarise the methods attachable to each component of the overall model. In each case they will correspond to a density evaluation method and a sampling method, but there are case-specific nuances. In addition to the standard components of the SSM, it is also possible to define a sequence of proposal distributions. 

These tables list the arguments that each function can accept. The intended usage and expected types of the data arguments are given in Table \ref{tab:data_cats}. In the tables \ref{tab:prior_methds}-\ref{tab:prop_methds} arguments marked with an asterisk (*) are always passed; others are optional and passed only if available in the data-set.

Following \pkg{PyTorch} convention, model parameters, $\bm\theta$, should be registered as class attributes of the Module and are not passed explicitly to functions.

\textbf{prior\_model:} The methods associated with the prior model, $p\bra{\x_{0}\mid \bm \theta}$, are detailed in Table \ref{tab:prior_methds}. The arguments \code{batch\_size} and \code{n_particles} are used to control the size of the sample drawn and correspond to $B$ and $K$ respectively. Below is an example of a Gaussian \code{prior\_model}:

\footnotesize
\begin{Code}
class GaussianPrior(pydpf.Module):
    def __init__(self, mean, cholesky_covariance, device, generator):
        super().__init__()
        self.mean = mean
        self.cholesky_covariance_ = cholesky_covariance
        self.device = device
        self.generator = generator

    def sample(self, batch_size, n_particles):
        standard_sample = torch.randn((batch_size, n_particles, self.mean.size(0)), 
            device=self.device, generator=self.generator)
        return self.mean + standard_sample @ self.cholesky_covariance.T

    # Constrain the cholesky_covariance to be lower-triangular with positive diagonal
    @pydpf.constrained_parameter
    def cholesky_covariance(self):
        tril = torch.tril(self.cholesky_covariance_)
        diag = tril.diagonal()
        diag.mul_(diag.sign())
        return self.cholesky_covariance_, tril
\end{Code}
\normalsize
\begin{table}[ht]
        \centering
        \begin{tabular}{|c|m{10em}|m{7em}|m{10em}|}
            \hline
             Alias&\centering Output&\centering Arguments&Required?\\
             \hline
             \hline
             \code{log\_density()}&The density at a given state under the prior. Tensor of size $(B\times K)$.&\makecell[l]{*state \\time \\control\\series\_metadata}&For non-bootstrap filtering.\\
             \hline
             \code{sample()}&A function to sample the prior. Tensor of size $(B\times K\times D_{s})$.&\makecell[l]{*batch\_size\\ *n\_particles\\time\\control\\series\_metadata}&For data-generation and bootstrap filtering.\\
             \hline
        \end{tabular}
        \caption{The functions defined for the \code{prior\_model} Module.}
        \label{tab:prior_methds}
    \end{table}
\textbf{dynamic\_model:} The methods associated with the dynamic model, $p\bra{\x_{t}\mid \x_{t-1}; \bm\theta}$, are described in Table \ref{tab:dynamic_methds}. We give the following example of a linear and Gaussian dynamic kernel:

\footnotesize
\begin{Code}
class LinearGaussianDynamic(pydpf.Module):
    def __init__(self, weight, bias, cholesky_covariance, device, generator, max_spectral_radius = 0.99):
        super().__init__()
        self.weight = weight
        self.bias = bias
        self.cholesky_covariance_ = cholesky_covariance
        self.device = device
        self.generator = generator
        self.max_spectral_radius = max_spectral_radius

    def sample(self, prev_state):
        standard_sample = torch.randn(prev_state.size(), device=self.device, generator=self.generator)
        mean = (self.constrained_weight @ prev_state.unsqueeze(-1)).squeeze() + self.bias
        return mean + standard_sample @ self.cholesky_covariance.T

    #Constrain the cholesky_covariance to be lower-triangular with positive diagonal
    @pydpf.constrained_parameter
    def cholesky_covariance(self):
        tril = torch.tril(self.cholesky_covariance_)
        diag = tril.diagonal()
        diag.mul_(diag.sign())
        return self.cholesky_covariance_, tril

    #Constrain the weight's spectral radius to avoid divergence
    @pydpf.constrained_parameter
    def constrained_weight(self):
        if self.max_spectral_radius is not None:
            eigvals = torch.linalg.eigvals(self.weight)
            spectral_radius = torch.max(torch.abs(eigvals))
            if spectral_radius > self.max_spectral_radius:
                return self.weight, self.weight / spectral_radius
        return self.weight, self.weight
\end{Code}
\normalsize
\begin{table}[ht]
        \centering
        \begin{tabular}{|c|m{10em}|m{7em}|m{10em}|}
            \hline
             Alias&\centering Usage&\centering Arguments&Required?\\
             \hline
             \hline
             \code{log\_density()}&The density at a given state under the dynamic kernel. Tensor of size $(B\times K)$.&\makecell[l]{*prev\_state\\ *state\\prev\_time\\time\\control\\series\_metadata\\ *t}&For non-bootstrap filtering.\\
             \hline
             \code{sample()}&A function to sample the dynamic kernel. Tensor of size $(B\times K\times D_{s})$.&\makecell[l]{*prev\_state\\prev\_time\\time\\control\\series\_metadata\\ *t}&For data-generation and bootstrap filtering.\\
             \hline
        \end{tabular}
        \caption{The functions defined for the \code{dynamic\_model} Module.}
        \label{tab:dynamic_methds}
\end{table}
\textbf{observation\_model:} The methods associated with the observation model, $p\bra{\y_{t}\mid \x_{t};\bm\theta}$, are described in Table \ref{tab:observation_methds}. We refer to the evaluation function related to the observation\_model as the score rather than the log density, this is because there is no requirement for the output to be a valid density. From this perspective, the observation model may be seen as a fitness function analogous to those used in genetic algorithms \citep{moral2004FK-formulae}. For example, several differentiable particle filtering applications have employed approximate Bayesian computation \citep{jonschkowski2018DPF, younis2024mixtureresample}. We now present an example of a linear and Gaussian observation model:

\footnotesize
\begin{Code}
class LinearGaussianObservation(pydpf.Module):
    def __init__(self, weight, bias, cholesky_covariance, device, generator):
        super().__init__()
        self.weight = weight
        self.bias = bias
        self.cholesky_covariance_ = cholesky_covariance
        self.device = device
        self.generator = generator

    def score(self, state, observation):
        mean = (self.weight @ state.unsqueeze(-1)).squeeze() + self.bias
        residuals = observation.unsqueeze(1) - mean
        exponent = (-1 / 2) * torch.sum((residuals @ self.inv_cholesky_covariance.T) ** 2, dim=-1)
        return self.density_pre_factor + exponent

    # Constrain the cholesky_covariance to be lower-triangular with positive diagonal
    @pydpf.constrained_parameter
    def cholesky_covariance(self):
        tril = torch.tril(self.cholesky_covariance_)
        diag = tril.diagonal()
        diag.mul_(diag.sign())
        return self.cholesky_covariance_, tril

    #Cache the inverse covariance to avoid recalculating it
    @pydpf.cached_property
    def inv_cholesky_covariance(self):
        return torch.linalg.inv_ex(self.cholesky_covariance)[0]

    #Cache the normalising constant for the density
    @pydpf.cached_property
    def density_pre_factor(self):
        return (-1/2 * self.weight.size(-1) * torch.log(torch.tensor(2*torch.pi)) 
            - torch.linalg.slogdet(self.cholesky_covariance)[1])
\end{Code}
    \begin{table}[ht]
        \small
        \centering
        \begin{tabular}{|c|m{10em}|m{7em}|m{10em}|}
            \hline
             Alias&\centering Usage&\centering Arguments&Required?\\
             \hline
             \hline
             \code{score()}&The score of an observation given the latent state. Tensor of size $(B\times K)$.&\makecell[l]{*state\\ *observation\\time\\control\\series\_metadata\\ *t}&For all filtering algorithms.\\
             \hline
             \code{sample()}&A function to sample from the observation kernel. Tensor of size $(B\times K \times D_{o})$.&\makecell[l]{*state\\prev\_time\\time\\control\\series\_metadata\\ *t}&For data-generation.\\
             \hline
        \end{tabular}
        \caption{The functions defined for the \code{observation\_model} Module.}
        \label{tab:observation_methds}
    \end{table}
\normalsize
\textbf{initial\_proposal\_model:} The methods associated with the initial proposal model, $\pi\bra{\x_{0}\mid\bm\theta}$ are described in Table \ref{tab:init_prop_methds}. We do not provide a code example for the \code{initial\_proposal\_model} as it is identical to the \code{prior_model} aside from the possibility to condition on the observation.

    \begin{table}[ht]
        \small
        \centering
        \begin{tabular}{|c|m{10em}|m{7em}|m{10em}|}
            \hline
             Alias&\centering Usage&\centering Arguments&Required?\\
             \hline
             \hline
             \code{log_density()}&The density at a given state under the initial proposal. Tensor of size $(B\times K)$.&\makecell[l]{*state\\ *observation\\prev\_time\\time\\control\\series\_metadata}&If using particle filters other than the bootstrap particle filter.\\
             \hline
             \code{sample()}&A function to sample from the initial proposal distribution. $(B\times K\times D_{s})$.&\makecell[l]{*batch\_size\\ *n\_particles\\ *observation\\time\\control\\series\_metadata}&If using particle filters other than the bootstrap particle filter.\\
             \hline
        \end{tabular}
        \caption{The functions defined for the \code{initial\_proposal\_model} Module.}
        \label{tab:init_prop_methds}
    \end{table}

\textbf{proposal\_model:} The methods associated with the proposal model, $\pi\bra{\x_{t}\mid \x_{t-1},\y_{t};\theta}$ are detailed in Table \ref{tab:prop_methds}. We do not provide a code example for the \code{proposal\_model} as it is identical to the \code{dynamic_model} aside from the possibility to condition on the observation.
\begin{table}[ht]
        \centering
        \begin{tabular}{|c|m{10em}|m{7em}|m{10em}|}
            \hline
             Alias&\centering Usage&\centering Arguments&Required?\\
             \hline
             \hline
             \code{log\_density()}&The density at a given state under the proposal kernel. Tensor of size $(B\times K)$.&\makecell[l]{*prev\_state\\ *state\\ *observation\\prev\_time\\time\\control\\series\_metadata\\ *t}&For non-bootstrap filtering.\\
             \hline
             \code{sample()}&A function to sample the proposal kernel. $(B\times K\times D_{s})$.&\makecell[l]{*prev\_state\\ *observation\\prev\_time\\time\\control\\series\_metadata\\ *t}&For non-bootstrap filtering.\\
             \hline
        \end{tabular}
        \caption{The functions defined for the \code{proposal\_model} Module.}
        \label{tab:prop_methds}
\end{table}

Finally having defined the model components, one can package them into a \code{pydpf.FilteringModel} object. This is as simple as passing the components as arguments to the constructor.

\footnotesize
\begin{Code}
#Define a FilteringModel from components
custom_model = FilteringModel(prior_model = custom_prior,
                              dynamic_model = custom_dynamic,
                              observation_model = custom_observation,
                              initial_proposal_model = custom_i_prop,
                              proposal_model = custom_prop)
\end{Code}
\normalsize
If the \code{initial_proposal_model} is not specified then the \code{prior_model} will be used in its place, and similarly with the \code{proposal_model} and \code{dynamic_model}. The resultant filter is known as the bootstrap filter \cite{Gordon93}. See Section \ref{sec:ssms:pf} for more details.

\subsubsection{Generating synthetic data in PyDPF}
Having defined an SSM with the required components, we can simulate trajectories from it and save them to a file in the format described in Section \ref{sec:basic:data}. We provide options to control the length of each trajectory, namely: \code{time_extent}; the total number of generated trajectories, \code{n_trajectories}; and the number of trajectories to generate at a time (using GPU parallelism if available), \code{batch_size}.

\footnotesize
\begin{Code}
pydpf.simulate_and_save(data_path, 
                        SSM=SSM, 
                        time_extent=1000, 
                        n_trajectories=2000, 
                        batch_size=100, 
                        device=device)
\end{Code}
\normalsize

\subsection{Particle filtering}
\label{sec:ssms:pf}
    A popular method to approximate the filtering distribution of a general SSM is the particle filter.
    The particle filter constructs a Monte Carlo approximation to the filtering distribution using importance sampling.
    A commonly used particle filtering algorithm is the sequential importance resampling (SIR) particle filter, which is given in Alg.~\ref{alg:bg:sir_pf}. 
    In this algorithm, we compute a set of weights and particles $\{(\x_t^{(k)}, w_t^{(k)})\}_{k=1}^{K}$ which gives a Monte Carlo estimate of the filtering distribution $p(\x_t | \y_{0:t})$ for each time $t$.
        \begin{algorithm}[ht]
            \caption{Sequential importance resampling (SIR) particle filter}
            \label{alg:bg:sir_pf}
            \begin{algorithmic}[1]
                \Input Observations $\y_{0:T}$, parameters $\bm\theta$.
                \Output Hidden state estimates $\x^{(1:K)}_{0:T}$, particle weights $w^{(1:K)}_{0:T}$.
                \State Sample $\x_0^{(k)} \sim p(\x_0|\bm\theta)$, for $k = 1,\dots,K$.
                \label{step:pf:prior_draw}
                \State Set $w_{0}^{(k)} = \frac{p(\y_0\mid\x^{(k)}_{0};\bm\theta)p(\x^{(k)}_0\mid\bm\theta)}{\pi_{0}\bra{\x^{(k)}_0 \mid y_{0};\bm\theta}}$, for $k=1,\dots,K$
                \State Set $\overline{w}_0^{(k)} = w^{(k)}_{0}/\sum^{K}_{k=1}w^{(k)}_{0}$, for $k = 1,\dots,K$.
                \For{$t = 1,\dots,T$}
                    \If{Resampling criterion} \label{step:pf:resample_crit}
                        \State Perform resampling with Alg.~\ref{alg:multi-res} to obtain $\widetilde{\x}_{t-1}^{(1:K)}$ and $\widetilde{w}_{t-1}^{(1:K)}$. \label{step:pf:resampling}
                    \Else{}
                        \State Set $\widetilde{\x}_{t-1}^{(k)} = {\x}_{t-1}^{(k)}$ and $\widetilde{w}_{t-1}^{(k)} = \overline{w}_{t-1}^{(k)}$. for $k = 1,\dots,K$. \label{step:pf:no_resampling}
                    \EndIf
                    \State Draw $\x_t^{(k)} \sim \pi(\x_t|\widetilde{\x}_{t-1}^{(k)}, \y_{t}; \bm\theta)$. for $k = 1,\dots,K$. \label{step:pf:draw}
                    
                    \State Set $w_t^{(k)} = \frac{p(\y_t|\x_t^{(k)}; \bm\theta)p(\x_t^{(k)}|\widetilde{\x}_{t-1}^{(k)}; \bm\theta)}{\pi(\x_t^{(k)}|\widetilde{\x}_{t-1}^{(k)}, \y_{t}; \bm\theta)}$. for $k = 1,\dots,K$.\label{step:pf:weights}
                    \State Set $\overline{w}_t^{(k)} = \widetilde{w}_{t-1}^{(k)}w_t^{(k)}/\sum_{k=1}^K\widetilde{w}_{t-1}^{(k)}w_t^{(k)}$. for $k = 1,\dots,K$.\label{step:pf:normweights}
                \EndFor
            \end{algorithmic}
        \end{algorithm}

        \begin{algorithm}[ht]
            \caption{Multinomial resampling}
            \label{alg:multi-res}
            \begin{algorithmic}[1]
                \Input Particles $\x^{(1:K)}_{t-1}$, normalised weights $\overline{w}^{(1:K)}_{t-1}$.
                \Output Resampled particles $\widetilde{\x}^{(1:K)}_{t-1}$, resampled weights $\widetilde{w}^{(1:K)}_{t-1}$.
                \For{$k = 1,\dots,K$}
                    \State Draw $a_{t}^{(k)} \sim \mathrm{Categorical}\bra{\overline{w}^{(1:K)}_{t-1}}$. \label{step:res:draw}
                    \State Set $\widetilde{w}_{t-1}^{(k)} = 1/K$. \label{step:res:resweights}
                    \State Set $\widetilde{\x}_{t-1}^{(k)} = \x^{\bra{a_{t}^{(k)}}}_{t-1}$.
                \EndFor
            \end{algorithmic}
        \end{algorithm}

    We explain the SIR particle filter below, following Alg.~\ref{alg:bg:sir_pf}.
    First, we initialise the particle set by drawing $K$ samples from a proposal distribution $\pi_{0}\bra{\x_0 \mid y_{0};\bm\theta}$, and setting the particle $\x_0^{(k)}$ equal to the $k^\text{th}$ sample for $k = 1, \dots, K$. We apply Bayes rule to weight the particles according to their posterior probability
    \begin{equation}
        w_0^{(k)} = \frac{p(\y_0\mid\x^{(k)}_{0};\bm\theta)p(\x^{(k)}_0\mid\bm\theta)}{\pi_{0}\bra{\x^{(k)}_0 \mid y_{0};\bm\theta}}, \quad \forall k=1,\dots,K.
    \end{equation}
    From this initialisation, the algorithm sequentially processes the remaining observation sequence $\y_{0:T}$, with the iteration at time-step $t$ proceeding as follows.
    
    First, we resample the particle set with replacement, drawing the $k^\text{th}$ index from a categorical distribution with event probabilities given by the normalised weights at time-step $t-1$, $\overline{w}_{t-1}$, which we write $a_{t}^{(k)} \sim \mathrm{Categorical}(\overline{w}_{t-1})$.
    This corresponds to line \ref{step:pf:resampling} of Alg.~\ref{alg:bg:sir_pf}.
    Note that this is equivalent to sampling from the multinomial distribution $\mathrm{Multinomial}(k, \overline{w}_{t-1})$.
    Resampling is vital to maintain the diversity of the particle set, and hence to obtaining accurate estimates of the filtering distribution $p(\x_t | \y_{0:t})$. 
    If resampling is not performed, then as the particle filter iterates, the weights are known to degenerate such that all but very few particle weights are close to $0$. This weight degeneracy renders the Monte Carlo approximation to the filtering distribution $p(\x_t | \y_{0:t})$ unusable \citep{doucet2009tutorial}.
    
    If we perform resampling, we set the historical normalised resampled weights $\widetilde{w}_{t-1}$ uniformly equal to $K^{-1}$.
    In many implementations of the particle filter, resampling is only performed if the effective sample size (ESS) of the previous particle weights $\overline{w}_{t-1}$, given by 
        \begin{equation}
            \label{eq:ess}
            \widehat{\text{ESS}}(\{w^{(k)}\}_{k=1}^{K}) = \frac{\left(\sum_{k=1}^{K}w^{(k)}\right)^2}{\sum_{k=1}^{K}\left(w^{(k)}\right)^2} \leq K,
        \end{equation}
    is less than some proportion of $K$, $pK, \ 0 < p \leq 1$.
    This is encoded in line \ref{step:pf:resample_crit}, where we perform resampling only if the resampling criterion is met.
    However, in a parallelised batch-sequential setting, evaluating the resampling criterion, e.g., the ESS introduces a performance overhead. So, it is most common in DPFs to perform resampling at every time-step. 

    After optionally resampling the particle set, we draw $K$ samples from the proposal distribution $\pi(\x_{t}| \x_{t-1}, \y_{t}; \bm\theta)$, following Line \ref{step:pf:draw} of Alg.~\ref{alg:bg:sir_pf}. 
    The proposal distribution is not uniquely defined by a state-space model.
    There are many choices for the proposal distribution, with some examples being the bootstrap particle filter \citep{Gordon93} and the auxiliary particle filter \citep{pitt1999filtering}.
    The bootstrap proposal, equivalent to choosing the proposal distribution to be the dynamic kernel, is particularly common as it allows the weight computation in Line \ref{step:pf:weights} of Alg. \ref{alg:bg:sir_pf} to be simplified.
    The proposal distribution is important to consider when designing a particle filter; a good proposal distribution should result in particles that are distributed across the state space roughly according to the posterior probability density.

    Next, we incorporate the current observation $\y_t$ via the importance weights, given by 
        \begin{equation}
            \label{eq:weight}
            w_t^{(k)} = \frac{p(\y_t|\x_t^{(k)}; \bm\theta)p(\x_t^{(k)}|\widetilde{\x}_{t-1}^{(k)}; \bm\theta)}{\pi(\x_t^{(k)}|\widetilde{\x}_{t-1}^{(k)}, \y_{t}; \bm\theta)}.
        \end{equation}
    We compute the weights at line \ref{step:pf:weights} of Alg.~\ref{alg:bg:sir_pf}.
    If using the bootstrap proposal \cite{Gordon93}, we have significant cancellation in the weight computation, Eq.~\eqref{eq:weight}, by noting that, for the bootstrap proposal, 
    \begin{equation}
        \pi_{\mathrm{bootstrap}}(\x_t^{(k)}|\widetilde{\x}_{t-1}^{(k)}, \y_{t}; \bm\theta) := p(\x_t^{(k)}|\widetilde{\x}_{t-1}^{(k)}; \bm\theta),
    \end{equation}
    and we therefore have 
    \begin{equation}
        \left(w_t^{(k)}\right)_{\mathrm{bootstrap}} = p(\y_t|\x_t^{(k)}; \bm\theta).
    \end{equation}
    This is particularly useful when the transition kernel $p(\x_t^{(k)}|\widetilde{\x}_{t-1}^{(k)}; \bm\theta)$ can be sampled but does not admit a tractable density.

    Finally, in line \ref{step:pf:normweights} of Alg.~\ref{alg:bg:sir_pf}, we normalise the weights by $\overline{w}_t^{(k)} = \widetilde{w}_{t-1}^{(k)}w_t^{(k)}/\sum_{k=1}^K\widetilde{w}_{t-1}^{(k)}w_t^{(k)}$, where we note that if we perform resampling at every step, we have $\widetilde{w}_{t-1}^{(k)} = K^{-1} \ \forall k, t$. 
    After performing weight normalisation for time-step $t$, the particle filter then proceeds to time-step $t+1$, where we repeat the above procedures.

    The particle filter consumes the entire observation series $\y_{0:T}$, and for each $\y_{t}, t = 0, \dots, T$, outputs particle-weight pairs, given by $\{(\x_t^{(k)}, w_t^{(k)})\}_{k=1}^{K}$.
    These can be used to construct importance estimates of expectations of the filtering distribution $p(\x_t | \y_{0:t})$ via
        \begin{equation}
            \label{eq:is_estimator}
            \expec_{p(\x_t | \y_{0:t})}\sbra{f(\x_t)} = \int f(\x_t) p(\x_t | \y_{0:t}) \mathrm{d}\x_t \approx \sum_{k=1}^K w_t^{(k)}f(\x_t^{(k)}).
        \end{equation}

\subsubsection{Running a particle filter in PyDPF}
\label{sec:ssms:pf:imp}
Having defined a state-space model and loaded the data it is simple to run a particle filter. In the example below we run a particle filter with $1000$ particles and multinomial resampling for every trajectory in the dataset. We pass the output of the filter through a function, labelled \code{aggregation_function}. This function can take all of the fields given in Table \ref{tab:data_cats}, and additionally the ground truth latent state, the particle weights, and the estimated likelihood factor $p\bra{y_{t}\mid y_{0:t-1}}$. The Tensor outputted from \code{aggregation_function} should not have a shape that depends on its inputs. We have implemented several common output functions and losses in \code{pydpf.outputs}. We introduce this function for memory efficiency, in most cases a well chosen \code{aggregation_function} can avoid having to store all variables generated in filtering. Note that this is most useful during inference; in training \pkg{PyTorch} retains many intermediates for the backwards pass. 

In \pkg{PyDPF}, filtering algorithms should be treated as models themselves and instantiated as \code{pydpf.Module} objects. This design pattern of treating algorithms as models is common in \pkg{PyTorch}, and in our case allows the flexibility to attach additional parameters to the algorithm that are not part of the SSM. Similarly for the \code{resampler}, this is useful for example to implement the algorithm of \cite{younis2024mixtureresample} in which the resampling algorithm depends on learned parameters. We provide a minimal code example for running a particle filter in \pkg{PyDPF}:

\footnotesize
\begin{Code}
multinomial_resampler = pydpf.MultinomialResampler(torch.Generator(device=device))
PF = pydpf.ParticleFilter(resampler=multinomial_resampler, SSM=SSM)
aggregation_function = pydpf.FilteringMean()
for state, observation in data_loader:
    estimated_state = PF(observation=observation, 
                         n_particles=1000, 
                         aggregation_function=aggregation_function, 
                         time_extent=1000)
\end{Code}
\normalsize

\subsection{Parameter estimation in state-space models}
\label{sec:param_infer}
    In order to utilise the particle filter described in Sec.~\ref{sec:ssms:pf}, we must know, or have suitable estimates for, the value of $\bm\theta$.
    However, in general $\bm\theta$ is unknown, and must be estimated. One typically obtains point estimates of $\bm\theta$ through maximising the joint likelihood of the observations, $p(\y_{0:T}\mid \bm\theta)$. The particle filtering estimate of which is given by:
    \begin{equation}
            \label{eq:likelihood_est}
            p(\y_{0:T} | \bm\theta) \approx \prod_{t=0}^{T} \left(\frac{1}{K}\sum_{k=1}^{K} \left(w_{t}^{(k)} \widetilde{w}_{t-1}^{(k)}\right)\right),
        \end{equation}
    with $w_{t}^{(k)}$ and $\widetilde{w}^{(k)}_{t-1}$ as per Alg.~\ref{alg:bg:sir_pf}. Since Alg.~\ref{alg:bg:sir_pf} involves random sampling it is not differentiable with respect to the parameters of the SSM, therefore direct optimisation of Eq.~\eqref{eq:likelihood_est} requires a scheme that is gradient free and robust to a noisy objective function. Methods such as Nelder-Mead \citep{nelder1965simplex} can be utilised in this instance, but are susceptible to local minima and require a large number of evaluations of the parameter likelihood, which is computationally expensive. First-order optimisation schemes such as Adam \citep{kingma2014adam} are known to converge in fewer iterations and be less susceptible to local minima than zeroth-order schemes such as Nelder-Mead. However, these schemes require gradient information.

    More popular in classical settings is Bayesian parameter estimation which targets the parameter posterior density, $p(\bm\theta | \y_{0:T})$, where we have
        \begin{equation}
            \label{eq:param_post}
            p(\bm\theta | \y_{0:T}) \propto p(\bm\theta) p(\y_{0:T} | \bm\theta).
        \end{equation}
    For example, particle MCMC \citep{andrieu2010particle}, wherein the posterior density $p(\bm\theta | \y_{0:T})$ is sampled using a standard MCMC scheme such as Metropolis-Hastings.
    It has been shown that, under very broad conditions, utilising the stochastic estimate for the parameter likelihood given in Eq.~\eqref{eq:likelihood_est} generates samples from the true posterior \citep{andrieu2010particle}.

    Other methods such as particle Gibbs \citep{andrieu2010particle} and particle Gibbs with ancestor sampling \citep{lindsten2014particle} build on this methodology, and generate sample-based approximations to the parameter posterior in the usual manner of MCMC methods.

    Of note is that all of these methods are gradient-free, as the parameter posterior density given in Eq.~\eqref{eq:param_post} is not differentiable with respect to the parameter values.
    Therefore, MCMC kernels such as Hamiltonian Monte Carlo  \citep{neal2011mcmc} and no u-turn Sampler  \citep{hoffman2014no} cannot be applied.

\section{Differentiable particle filters}
    \label{sec:dpfs}
    The general SIR particle filter described in Alg.~\ref{alg:bg:sir_pf} and Section \ref{sec:ssms:pf} is not differentiable, as it requires drawing samples from a discrete distribution.
    
    In particular, sampling the categorical and multinomial distribution depends on a series of real-valued probabilities, for which an infinitesimal change in value can yield a discrete change in the sample value, thereby rendering a direct sampling procedure non-differentiable.

    

\subsection{Monte Carlo gradient estimation}
    \label{sec:bg:MCgrad}
    Broadly, a differentiable particle filter (DPF) is an algorithm that simultaneously returns Monte Carlo estimators of both expectations with respect to the posterior of functions of the latent state and their gradient.
    If $\x$ is a sample from a probability distribution $p(\x; \bm\theta)$ on $\reals^{d_{x}}$ that depends explicitly on parameters $\bm\theta$, then the gradient of $\x$ with respect to $\bm \theta$, $\grad\x$, is not defined.
    But, the gradient of its expectation, $\grad\expec_{{p(\x; \bm\theta)}} \sbra{\x}$ is. Typically it is analytically intractable, so we approximate it via a Monte Carlo estimator.
    This section briefly outlines important methods for Monte Carlo gradient estimation, for an in-depth overview we refer the reader to \citet{mohamed2020gradientestimation}. 
    Throughout this section we assume that the regularity conditions that allow the interchange of the differentiation and integration operators are always satisfied.

\subsubsection{Reparametrisation trick}
    The reparametrisation trick \citep{kingma2013VAE} applies when $\x \sim p(\x; \bm\theta)$ may be generated as a differentiable transformation of a sample from an auxiliary distribution that does not depend on $\bm \theta$, i.e. taking $\x = f\bra{\z; \bm \theta}, \; \z \sim q(\z)$ simulates $\x \sim p(\x; \bm\theta)$.
    Having sampled $\x$ using the reparametrisation trick the gradient may be sampled by vanilla back-propagation, $\z$ has no dependence on $\bm \theta$ so $\grad \z = 0$.
    It is trivial to show that the resultant gradient estimator is unbiased.
    
    An example of a distribution that admits the reparametrisation trick is the multivariate Gaussian distribution. 
    Let $\x \sim \mathcal{N}(\bm \mu, \mathbf{S}\mathbf{S}^{T})$, then we have  $\x \simeq \bm\mu + \mathbf{S}\bm\epsilon$, where $\bm\epsilon \sim \mathcal{N}(\bm 0, \mathbf{Id})$.
    As $\bm \epsilon$ is independent of $\bm\mu$ and $\mathbf{S}$, we can easily compute the gradient of $\x$ with respect to $\bm\mu$ and $\mathbf{S}$.

    The reparametrisation trick is low variance, computationally cheap and easy to implement, as such it is the default choice when a suitable function $f$ is available. The reparametrisation trick forms the basis of several popular deep sampling architectures, including the variational auto-encoder \citep{kingma2013VAE}, and normalising flows \citep{papamakarios2021normalizing}.

\subsubsection{REINFORCE}
    The REINFORCE estimator \citep{williams1992REINFORCE}, also known as the score function estimator or the likelihood ratio estimator, is a more generally applicable gradient estimator for sampling than the reparameterisation trick.
    REINFORCE requires that we are able to sample from the distribution and evaluate its probability density function.
    Let $\x \sim p(\x; \bm\theta)$, then:
        \begin{equation}
        \label{eq:REINFORCE}
            \begin{aligned}
            \grad\expec_{p(\x; \bm\theta)}\sbra{\psi\bra{\x}} &= \grad \int_{\reals^{d_{x}}}\psi\bra{\x}p\bra{\x;\bm\theta}d\x \\
            &= \int_{\reals^{d_{x}}}\psi\bra{\x}\frac{\grad p\bra{\x;\bm\theta}}{p\bra{\x;\theta}}p\bra{\x;\bm\theta}d\x \\
            &= \expec_{p(\x; \bm\theta)}\sbra{\psi\bra{\x}\grad \log p\bra{\x;\bm\theta}}
            \end{aligned}
        \end{equation}
    for some sufficiently regular test-function $\psi: \reals^{d_{x}} \to \reals$ independent of $\bm \theta$.
    Eq. \eqref{eq:REINFORCE} directly yields the appropriate gradient estimator, with the gradient $\grad \log p\bra{\x;\bm\theta}$ typically obtained by auto-differentiation.
    Furthermore, it is simple to extend REINFORCE to discrete random variables by replacing the integral with a sum over the appropriate domain.
    In practice we treat the gradient term in Eq. \eqref{eq:REINFORCE} like an importance weight as detailed in \cite{foerster2018Dice}, this is computationally efficient and makes the generalisation to importance weighted estimators clear.

    REINFORCE frequently suffers from high variance, and it is therefore common to use some form of variance reduction, see e.g. \citet{paisley2012controlvariates}.
    Our recommendation is to use REINFORCE only when an appropriate reparametrisation of the proposal distribution is not available.

\subsubsection{Application to particle filtering}
    In the SIR particle filter's (Alg. \ref{alg:bg:sir_pf}) main loop there are two sampling operations: drawing the particles from the proposal; and resampling. Most current work in differentiable particle filter considers proposal models that admit a reparametrisation \citep{jonschkowski2018DPF, karkus2018DPF, Corenflos2021OT, younis2024mixtureresample, chen2024normalisingflow}.
    However, vanilla resampling is discrete and no smooth reparametrisation exists, one is forced to either use REINFORCE \citep{scibior2021stopgrad}, modify the resampling step \citep{Corenflos2021OT}, or to ignore gradient terms \citep{jonschkowski2018DPF}.

    Differentiable particle filters (DPFs) refer to particle filters that define a gradient with respect to their outputs with respect to the parameters of the SSM and/or proposal model. We will now describe several DPF methods that are implemented in \pkg{PyDPF}.
    
\section{Implemented algorithms}
\label{sec:imp-alg}
For illustration, in this section we assume that the proposal distribution is reparameterised as this is by far the more common case in the DPF literature. However, \pkg{PyDPF} has implementations for the algorithms in \cite{scibior2021stopgrad} where the proposal distribution is not reparameterised; and can easily be extended to settings where the proposal is reparameterised for only some of the state dimensions such as in \cite{brady2025DIMMPF}. We categorise DPFs by how they propagate gradient through the resampling step; specific SSM architectures are not implemented in \pkg{PyDPF}.
\subsection{Non-differentiable resampling}
\label{sec:imp-alg:DPF}
In the algorithm of \cite{jonschkowski2018DPF}, gradients are not passed through resampling. At each time-step the derivatives of the outputs of resampling (the resampled states and weights) with respect to the inputs to resampling (the states and weights at the previous time-step) are set to zero. Therefore gradients are not accumulated over time-steps, so this algorithm can be seen as using a form of truncated back-propagation through time where the gradient is truncated at every time-step. Consequently, it produces gradient estimates with a low variance but high bias compared to other algorithms implemented in \pkg{PyDPF}.

\subsubsection{Defining a particle filter with non-differentiable resampling in PyDPF}
    Whilst the DPF with non-differentiable resampling, or indeed any DPF implemented in \pkg{PyDPF}, can be constructed from a base filtering algorithm and the relevant resampler, as demonstrated in Section \ref{sec:ssms:pf}, we provide convenience functions for all DPFs packaged in \pkg{PyDPF}. To instantiate a DPF with non-differentiable resampling one can call:
    
    \footnotesize
    \begin{Code}
        dpf = pydpf.DPF(SSM=SSM, resampling_generator=generator, multinomial=False)
    \end{Code}
    \normalsize
    \code{resampling_generator} is the \code{torch.Generator} object that will track the random state used during resampling, \code{multinomial} will perform resampling with the standard multinomial resampler if \code{True} otherwise it uses the systematic resampler of \cite{carpenter1999systematic}.

\subsubsection{Summary}
\begin{itemize}
        \item advantages
            \begin{itemize}
                \item Fast.
                \item Identical in the forward pass to SIR particle filtering, Algorithm \ref{alg:bg:sir_pf}.
                \item Comparatively low variance.
            \end{itemize}
        \item disadvantages
            \begin{itemize}
                \item High bias.
                \item Gradient information is not propagated through time-steps.
            \end{itemize}
    \end{itemize}

\subsection{Soft resampling}
\label{sec:imp:soft}
    \citet{karkus2018DPF} modifies the resampling procedure such that the weights of the resampled particles depend on the weights of the pre-resampling particles in a differentiable way.
    The resampling step of the SIR particle filter (line \ref{step:pf:resampling} of Alg.~\ref{alg:bg:sir_pf} is replaced with a call to Alg.~\ref{alg:soft_res}).

        \begin{algorithm}[ht]
            \caption{Soft resampling}
            \label{alg:soft_res}
            \begin{algorithmic}[1]
                \Input Particles $\{\x_{t-1}^{(k)}\}^{K}_{k=1}$, normalised weights $\{\overline{w}^{(k)}_{t-1}\}^{K}_{k=1}$.
                \Output Resampled particles $\{\widetilde{\x}_{t-1}^{(k)}\}^{K}_{k=1}$, resampled weights $\{\widetilde{w}^{(k)}_{t-1}\}^{K}_{k=1}$.
                \For{$k = 1,\dots,K$}
                    \State Set $\bar{w}'^{\bra{k}}_{t-1} = \xi \bar{w}^{\bra{k}}_{t-1} + \frac{\bra{1-\xi}}{K}$
                    \State Draw $a_{t}^{(k)} \sim \mathrm{Categorical}(\overline{w}'_{t-1})$. \label{step:soft_res:draw}
                    \State Set $\tilde{w}^{\bra{k}}_{t-1} = \frac{\bar{w}^{\bra{a^{\bra{k}}_{t}}}_{t-1}}{K\bar{w}'^{\bra{a^{\bra{k}}_{t}}}_{t-1}} \; $. \label{step:soft_res:resweights}
                    \vspace{0.1em}
                    \State Set $\widetilde{\x}_{t-1}^{(k)} = \x^{\bra{a_{t}^{(k)}}}_{t-1}$.
                \EndFor
            \end{algorithmic}
        \end{algorithm}
        
    Each particle is assigned resampling weight $\xi \bar{w}^{\bra{k}}_{t-1} + \frac{\bra{1-\xi}}{K}; \; \xi \in [0, 1]$, where $\xi$ is a hyper-parameter and $K$ is the number of particles, so that resampling induces the importance weight,
        \begin{equation}
        \label{eq:if:softweights}
            \tilde{w}^{\bra{k}}_{t-1} = \frac{\bar{w}^{\bra{a^{\bra{k}}_{t}}}_{t-1}}{K\bar{w}'^{\bra{k}}_{t-1}} \; .
        \end{equation}
    Eq. \eqref{eq:if:softweights} is partially differentiable, gradients are taken with respect to $\bar{w}^{\bra{a^{\bra{k}}_{t}}}_{t-1}$ but not $a^{\bra{k}}_{t}$, so soft resampling returns biased gradient estimates.
    
    When $\xi = 1$, the resampling distribution is unmodified from usual resampling and no gradient is passed to the new weights, this is the strategy employed in \cite{le2018auto-encoding, maddison2017variational}, and with $\xi = 0$ particles are chosen uniformly and $a^{\bra{k}}_{t}$ is independent of the model parameters.
    Soft-resampling can be thought of as trading off between statistically efficient sampling and unbiased gradient estimation.
    Unlike the non-differentiable resampling described in Section \ref{sec:imp-alg:DPF}, soft-resampling carries forward the gradient of the particles between time-steps.

\subsubsection{Defining a particle filter with soft resampling in PyDPF}
\label{sec:imp:soft:def}
    To instantiate a DPF with soft-resampling one can call:
    
    \footnotesize
    \begin{Code}
        dpf = pydpf.SoftDPF(SSM=SSM, resampling_generator=generator, multinomial=False)
    \end{Code}
    \normalsize
    Where \code{resampling_generator} is the \code{torch.Generator} object that will track the random state used during resampling, \code{multinomial} will perform resampling with the standard multinomial resampler if \code{True} otherwise it uses the systematic resampler of \cite{carpenter1999systematic}.
    
\subsubsection{Summary}
\begin{itemize}
        \item advantages
            \begin{itemize}
                \item Relatively fast.
                \item Consistent forward pass.
                \item Flexibility to tune $\xi$ for the optimal bias-variance trade-off.
            \end{itemize}
        \item disadvantages
            \begin{itemize}
                \item Non-consistent backwards pass when $\xi \neq 0$ and unacceptably variant on both the forward and backward passes when $\xi = 0$.
                \item Requires the tuning of an extra hyper-parameter.
            \end{itemize}
    \end{itemize}

\subsection{Optimal transport resampling}
        \begin{algorithm}[ht]
            \caption{Sinkhorn Algorithm \hfill $\mathrm{Sinkhorn}(\w, \v, \X, \Y, \epsilon)$. \null}
            \label{alg:sinkhorn}
            \begin{algorithmic}[1]
                \Input Weight vectors $\w, \v$, sample matrices $\X, \Y$, regularisation strength $\epsilon$.
                \Output Transport vectors $\f, \g$, distance matrix $\C$.
                \State Initialise $\f = \bm 0, \g = \bm 0.$
                \State Set $\C = \X\X^{T} + \Y\Y^{T} - 2\X\Y^{T}.$
                \While{stopping criterion not met}
                    \For{$n = 1, \dots, K$}
                        \State Set $f_n = \frac{1}{2}\left(f_n + -\epsilon \, \mathrm{logsumexp}\left(\log(\v) + \epsilon^{-1}(\g - \C_{n,\cdot})\right)\right)$.
                        \State Set $g_n = \frac{1}{2}\left(g_n + -\epsilon \, \mathrm{logsumexp}\left(\log(\w) + \epsilon^{-1}(\f - \C_{\cdot,n})\right)\right)$.
                    \EndFor
                \EndWhile
            \end{algorithmic}
            where $\C_{n,\cdot}$ (resp. $\C_{\cdot,n}$) is the $n$th row (resp. column) of $\C$.
        \end{algorithm}

        \begin{algorithm}[ht]
            \caption{Optimal transport resampling}
            \label{alg:ot_res}
            \begin{algorithmic}[1]
                \Input Particles $\{\x_{t-1}^{(k)}\}^{K}_{k=1}$, normalised weights $\{\overline{w}^{(k)}_{t-1}\}^{K}_{k=1}$.
                \Output Resampled particles $\{\widetilde{\x}_{t-1}^{(k)}\}^{K}_{k=1}$, resampled weights $\{\widetilde{w}^{(k)}_{t-1}\}^{K}_{k=1}$.
                \State Set $\X$ such that $\X_{k,\cdot} = \x_{t-1}^{(k)} \ \forall k \in 1, \dots, K$.
                \State Set $\overline{\w}_{t-1}$ such that $(\overline{\w}_{t-1})_k = \overline{w}^{(k)}_{t-1} \ \forall k \in 1, \dots, K$.
                 \State Set $(\f, \g, \C) = \mathrm{Sinkhorn}(\overline{\w}_{t-1}, \bm 1^{(k)}/K, \X, \X)$. (Alg. \ref{alg:sinkhorn})
                \For{$n = 1,\dots,K$}
                    \For{$m = 1, \dots, K$}
                        \State Set $P_{n, m}^{(\epsilon)} = \frac{\overline{w}^{(n)}_{t-1}}{K} \exp\left(\frac{f_n + g_m - C_{n,m}}{\epsilon}\right).$ \label{step:ot:transport_matrix}
                    \EndFor
                \EndFor
                \State Set $\widetilde{\X} = K \mathbf{P}^{(\epsilon)}\X$.
                \For{$k = 1, \dots, K$}
                    \State Set $\widetilde{\x}_{t-1}^{(k)} = \widetilde{\X}_{k, \cdot}$.
                    \State Set $\widetilde{w}^{(k)}_{t-1} = 1/K$.
                \EndFor
            \end{algorithmic}
            where $\bm 1^{(k)}$ is a $k$-vector with every element equal to $1$.
        \end{algorithm}

    Optimal transport resampling \citep{Corenflos2021OT} replaces the stochastic resampling of the SIR particle filter with a deterministic and differentiable transport map. 
    This is performed by replacing line \ref{step:pf:resampling} of Alg.~\ref{alg:bg:sir_pf} with Alg.~\ref{alg:ot_res}. Our implementation of the Sinkhorn loop in Alg.~\ref{alg:sinkhorn} is a \pkg{PyTorch} reimplementation of \pkg{FilterFlow}'s \citep{Corenflos2021OT}. We decay the regularisation strength, $\epsilon$, over iterations from the diameter of the bounding sphere of the particle states after they have been scaled to have a standard deviation of one along each dimension. The decay is stopped once $\epsilon$ reaches a specified minimum. The specific stopping criterion we adopt is to halt the loop when either the algorithm has run for a specified maximum number of iterations, or both the following criteria are met: $\epsilon$ has reached its specified minimum and the update to the potentials is below a specified threshold.
    
    Formally, the map is from an empirical sample of the proposal distribution, \\$\int_{\reals^{d_x}}p\bra{\x_{t-1}|\y_{0:t-1}; \bm\theta}\pi\bra{\x_{t} | \x_{0:t-1}, \y_{t}; \bm\theta} d\x_{t-1}  \approx \frac{1}{K}\sum^{K}_{k=1} \delta^{d_x}\bra{\x_{t} - \x^{\bra{k}}_{t}}$, to the weighted posterior, $p\bra{\x_{t} | \y_{0:t}; \bm \theta} \approx \sum^{K}_{k=1} \bar{w}^{\bra{k}}_{t} \delta^{d_x}\bra{\x_{t} - \x^{\bra{k}}_{t}}$.
    They represent the transport map $T_{t-1}$, where $T^{\bra{ij}}$ is the weight from the $i^{\text{th}}$ particle to re-assign to the $j^{\text{th}}$ particle.
    The new particles are given by:
        \begin{subequations}
        \label{eq:if:otresample}
            \begin{gather}
                \tilde{\x}^{\bra{j}}_{t-1} = \sum^{K}_{i=1}\x^{\bra{i}}_{t-1}T^{\bra{ij}}\; , \\
                \tilde{w}^{\bra{k}}_{t-1} = \frac{1}{K} \; .
            \end{gather}
        \end{subequations}
    Any valid such map has $\sum^{K}_{i=1}T^{\bra{ij}} = \bar{w}^{\bra{j}}_{t}$, $\sum^{K}_{j=1}T^{\bra{ij}} = \frac{1}{K}$.
    The chosen map is the entropy regularised 2-Wasserstein optimal map \citep{cuturi2013sinkhorn}.
    \citet{Corenflos2021OT} prove that the resulting filter provides statistically consistent estimates of expectations of functions of the latent state and their gradients with respect to the model parameters.
    However, the application of Eq.~\eqref{eq:if:otresample} places the particles at new positions.
    This has three potentially problematic consequences: optimal transport resampling is not straightforwardly applicable if a component of the state space is discrete; it has increased sensitivity to the sharpness of posterior modes; and the likelihood estimates returned by a filter with optimal transport resampling are biased.

    Additionally, and in practice most importantly, optimal transport resampling suffers from high computational cost with $\mathcal{O}\bra{K^{2}}$ work and memory costs.
    In particular, every iteration of the loop in the Sinkhorn algorithm requires two calls to \pkg{PyTorch}'s costly \code{logsumexp} method.

    \subsubsection{Hyper-parameters}
    Our implementation of Algorithm \ref{alg:ot_res} closely follows \texttt{FilterFlow} \citep{Corenflos2021OT}. We follow their implementation by including a number of additional hyper-parameters that can be tuned to balance Monte Carlo bias, gradient variance, numerical stability, and execution time.
    
    In principle, $\epsilon$ can be learned but we treat it as a hyperparameter in all our experiments. We recommend scaling $\epsilon \propto \frac{1}{\log K}$ as this guarantees the Monte Carlo error vanishes as $K \to \infty$ \citep{chen2024normalisingflow}, but the choice of an appropriate absolute value is specific to the SSM.

    The non-regularised Kantorovich transport matrix is the solution of a linear programming problem where the objective function, but not the constraints depend on the cost function. Therefore the limiting case as $\epsilon \to 0$ has the transport matrix as a non-Lipschitz function of the cost. But low $\epsilon$ is desirable as Monte Carlo error of Alg. \ref{alg:bg:sir_pf} with Alg. \ref{alg:ot_res} scales as $\mathcal{O}\bra{\sqrt{\epsilon}}$ assuming all other parameters and hyper-parameters are held constant.

    To improve stability, we decay the regularisation strength from the maximum of $\epsilon$ and the diameter of the particle state after being normalised to mean zero and standard deviation one. The parameter \code{decay_rate} controls this behaviour.

    If, during the Sinkhorn loop, the potentials are modified by less than the hyper-parameter \code{min_update_size} for all batches the algorithm is considered converged and stopped early.

    The hyperparameter \code{max_iterations} is the maximum number of Sinkhorn iterations to run, regardless of convergence.

    The hyperparameter \code{transport_gradient_clip} is the value to clip each element of the gradient vector of the loss with respect to the transport matrix at, this is set by default to $1.0$ in \cite{Corenflos2021OT} but we find in our experiments that it has little effect on stability.

\subsubsection{Defining a particle filter with optimal transport resampling in PyDPF}
To instantiate a DPF with optimal transport resampling one can call:

    \footnotesize
    \begin{Code}
        dpf = pydpf.OptimalTransportDPF(SSM = SSM, 
                                        regularisation = 0.1, 
                                        decay_rate = 0.9, 
                                        min_update_size = 0.01,
                                        max_iterations = 100,
                                        transport_gradient_clip = 1.)
    \end{Code}
    \normalsize
    Where the usage of the arguments is detailed in the previous section.
    
\subsubsection{Summary}
    \begin{itemize}
        \item advantages
            \begin{itemize}
                \item Unbiased and consistent backwards pass under the filtering algorithm, as the number of particles approaches $\infty$.
                \item Consistent forwards pass, as the number of particles approaches $\infty$ and the regularisation parameter $\epsilon$ approaches $0$.
            \end{itemize}
        \item disadvantages
            \begin{itemize}
                \item High variance of the gradients.
                \item Extremely slow.
                \item Requires the tuning of additional hyper-parameters.
            \end{itemize}
    \end{itemize}

\subsection{Stop-gradient resampling}

                    

        \begin{algorithm}[ht]
            \caption{Stop gradient resampling}
            \label{alg:stop_grad_res}
            \begin{algorithmic}[1]
                \Input Hidden state estimates $\{\x_{t-1}^{(k)}\}^{K}_{k=1}$, normalised weights $\{\bar{w}^{(k)}_{t-1}\}^{K}_{k=1}$.
                \Output Particles $\{\widetilde{\x}_{t-1}^{(k)}\}^{K}_{k=1}$, resampled weights $\{\widetilde{w}^{(k)}_{t-1}\}^{K}_{k=1}$.
                \For{$k = 1,\dots,K$}
                    \State Draw $a_{t}^{(k)} \sim \mathrm{Categorical}(\bot\left(\overline{w}_{t-1})\right)$. \label{step:stop_grad:draw}
                    \State Set $\tilde{w}^{\bra{k}}_{t-1} = \frac{\bar{w}^{\bra{a^{\bra{k}}_{t}}}_{t-1}}{K \bot\sbra{\bar{w}^{\bra{a^{\bra{k}}_{t}}}_{t-1}}}$.\label{step:stop_grad:resweights}
                    \State Set $\widetilde{\x}_{t-1}^{(k)} = \x^{\bra{a_{t}^{(k)}}}_{t-1}$
                \EndFor
            \end{algorithmic}
        \end{algorithm}

    Stop-gradient resampling \citep{scibior2021stopgrad} provides gradient estimates without modifying the filtering estimates.
    Particle filters with stop-gradient resampling use REINFORCE to obtain estimates of the gradient of the loss.
    They replace the resampling step (line \ref{step:pf:resampling}) in the SIR algorithm (Alg.~\ref{alg:bg:sir_pf}) with Alg.~\ref{alg:stop_grad_res}, with the key change being
        \begin{equation}
        \label{eq:if:stopgrad}
            \text{Set }\tilde{w}^{\bra{k}}_{t-1} = \frac{\bar{w}^{\bra{a^{\bra{k}}_{t}}}_{t-1}}{K \bot\sbra{\bar{w}^{\bra{a^{\bra{k}}_{t}}}_{t-1}}} \, ,
        \end{equation}
    where $\bot\sbra{\cdot}$ is the `stop-gradient operator', defined as the operator that returns the enclosed quantity during the forward pass, but sets its gradient to zero during auto-differentiation.
    Comparison with Eq.~\eqref{eq:REINFORCE} shows that auto-differentiation with the gradient modified in this way returns the usual REINFORCE estimator for the gradient with respect to $\bar{w}^{\bra{0:K}}_{t-1}$ for all per-particle losses with the form $\mathcal{L}^{\bra{k}}_{t-1} = \psi\bra{\x_{t-1}^{\bra{a^{\bra{k}}_{t}}}}\tilde{w}^{\bra{k}}_{t-1}$.

    The analysis in \citet{scibior2021stopgrad} further shows that auto-differentiating through the complete computation graph resulting from Alg.~\ref{alg:bg:sir_pf} with Eq. \eqref{eq:if:stopgrad} leads to a statistically consistent gradient estimator for this class of loss functions, including filtering means and the evidence lower bound (ELBO).
    
    Unfortunately, stop-gradient resampling with Eq.~\eqref{eq:if:stopgrad} can lead to high variance.
    \citet{scibior2021stopgrad} additionally propose a stabilised variant at the cost of requiring $\mathcal{O}\bra{K^{2}}$ computational effort. They follow the weighting strategy used in marginal particle filters, developed independently in \citet{Klaas2005marginal} and \citet{elvira2019auxiliaryMIS}.
    Instead of replacing step \ref{step:pf:resampling} with Alg.~\ref{alg:stop_grad_res}, they replace step \ref{step:pf:weights} of Alg.~\ref{alg:bg:sir_pf} with

        \begin{equation}
        \label{eq:if:stop-grad-marginal}
            \text{Set } w^{\bra{k}}_{t} = p\bra{\y_{t}|\x\upindex{k}_{t}; \bm \theta}\frac{\sum^{K}_{i=1}\bar{w}\upindex{i}_{t-1}p\bra{\x\upindex{k}_{t}|\x\upindex{i}_{t-1}; \bm \theta}}{\sum^{K}_{i=1}\bot\sbra{\bar{w}\upindex{i}_{t-1}}\pi\bra{\x\upindex{k}_{t}|\x\upindex{i}_{t-1}; \bm \theta}} \,.
        \end{equation}
    Eq. \eqref{eq:if:stop-grad-marginal} provides a consistent estimator of the gradient of the ELBO and filtering means with respect to the parameters of the dynamic kernel $p$ if, during back-propagation, the gradient of $\cbra{\x^{(k)}_{t}}^{K,T}_{k=1,t=0}$ with respect to the parameters of the dynamic kernel is set to $0$.
    Eq. \eqref{eq:if:stop-grad-marginal} also provides a consistent estimator of the gradient of the ELBO and filtering means with respect to the parameters of the dynamic kernel $p$ for a bootstrap filter if the proposal is reparameterised \citep{brady2025DIMMPF}. It cannot, however, provide a principled estimator for gradients taken with respect to the parameters of a non-bootstrap proposal distribution.
        
    The marginal particle filter based estimator takes into account that a given particle can have been, in principle, be resampled from any ancestor at the previous time-step.
    The operation and space complexity of computing this estimator are $\mathcal{O}\bra{K^{2}}$, but for most practical choices of $p$ and $\pi$ it will be significantly faster to compute than optimal transport resampling since these operations can be executed in parallel and does not entail as many costly \code{logsumexp} calls.


\subsubsection{Defining a particle filter with stop-gradient resampling in PyDPF}
To instantiate a DPF with stop-gradient resampling, Eq. \eqref{eq:if:stopgrad} one can call:

    \footnotesize
    \begin{Code}
        dpf = pydpf.StopGradientDPF(SSM=SSM, 
                                    resampling_generator=generator, 
                                    multinomial=False)
    \end{Code}
    \normalsize
and instantiate a marginal stop-gradient DPF:

    \footnotesize
    \begin{Code}
        dpf = pydpf.MarginalStopGradientDPF(SSM=SSM, 
                                            resampling_generator=generator, 
                                            multinomial=False)
    \end{Code}
    \normalsize

    In both cases, \code{resampling_generator} is the \code{torch.Generator} object that will track the random state used during resampling, \code{multinomial} will perform resampling with the standard multinomial resampler if \code{True} otherwise it uses the systematic resampler of \cite{carpenter1999systematic}.

\subsubsection{Summary}

    \begin{itemize}
        \item advantages
            \begin{itemize}
                \item Consistent backwards pass under the true filtering distribution for the model parameters, including when the model is used as the proposal, \emph{i.e.} bootstrap filtering.
                \item Has a lower variance but more computationally costly variant, Eqs. \eqref{eq:if:stop-grad-marginal} if the problem requires it.
                \item Identical in the forward pass to SIR particle filtering, Algorithm \ref{alg:bg:sir_pf}, if using \eqref{eq:if:stopgrad} or the marginal particle filter \citep{Klaas2005marginal} if using \eqref{eq:if:stop-grad-marginal}.
            \end{itemize}
        \item disadvantages
            \begin{itemize}
                \item Asymptotically biased gradients for the parameters of the proposal distribution when not using the bootstrap proposal.
                \item Non-asymptotically biased gradients.
            \end{itemize}
    \end{itemize}

\subsection{Kernel mixture resampling}

        \begin{algorithm}[ht]
            \caption{Kernel-mixture resampling}
            \label{alg:kernel_res}
            \begin{algorithmic}[1]
                \Input Hidden state estimates $\{\x_{t-1}^{(k)}\}^{K}_{k=1}$, normalised weights $\{\bar{w}^{(k)}_{t-1}\}^{K}_{k=1}$, zero-centred symmetric kernel with density $\phi\bra{\cdot}$.
                \Output Particles $\{\widetilde{\x}_{t-1}^{(k)}\}^{K}_{k=1}$, resampled weights $\{\widetilde{w}^{(k)}_{t-1}\}^{K}_{k=1}$.
                \For{$k = 1,\dots,K$}
                    \State Draw $\widetilde{\x}_{t-1}^{(k)} \sim \sum^{K}_{l=1}\bar{w}^{(l)}_{t-1}\phi\bra{\widetilde{\x}_{t-1}^{(k)} - \x^{(l)}_{t-1}}$. \label{step:kernel:draw}
                    \State Set $\tilde{w}^{\bra{k}}_{t-1} = \frac{\sum^{K}_{l=1}\bar{w}^{(l)}_{t-1}\phi\bra{\widetilde{\x}_{t-1}^{(k)} - \x^{(l)}_{t-1}}}{K \bot\sbra{\sum^{K}_{l=1}\bar{w}^{(l)}_{t-1}\phi\bra{\widetilde{\x}_{t-1}^{(k)} - \x^{(l)}_{t-1}}}}$.\label{step:kernel:resweights}
                \EndFor
            \end{algorithmic}
        \end{algorithm}
    \citet{younis2024mixtureresample} proposed a resampling method based on the post-regularised particle filter \citep{Musso2001RegPF}.
    Instead of multinomial resampling, the particles are drawn from a kernel density estimator with the symmetric kernels centred at the particle locations.
    The gradient due to sampling is estimated using REINFORCE.
    Like marginal stop-gradient resampling, the gradient of the new particles depend on the entire population of particles at the previous time-step, thereby lessening the variance induced by path-degeneracy.

    Kernel resampling is motivated by the intuition that applying Kernel smoothing to the particle field may help to stabilise the gradients. However, the gradients generated by the Kernel-mixture resampling particle filter enjoy less theoretical support than with the stop-gradient and optimal transport resamplers.

\subsubsection{Defining a particle filter with kernel resampling in PyDPF}
Instantiating a particle filter with kernel resampling is more complicated than the other DPFs due to needing to define the kernel. In this example we assume that the hidden state has dimension one and use a uni-dimensional Gaussian kernel. We initialise the kernel with zero mean and unit variance but allow the variance to be learned.

\footnotesize
\begin{Code}
    kernel = pydpf.StandardGaussian(dim = 1, 
                                    generator = generator, 
                                    learn_mean = False, 
                                    learn_cov = True)
    kernel_mixture = pydpf.KernelMixture(kernel, generator = generator)
    dpf = pydpf.KernelDPF(SSM=SSM, kernel = kernel_mixture)
\end{Code}
\normalsize

\subsubsection{Summary}
    \begin{itemize}
        \item advantages
            \begin{itemize}
                \item The gradient of resampled particles depends on the entire population of particles at the previous time-step.
                \item Regularised particle filters are rigorously theoretically supported \citep{leGland1998regularised-pf-analysis}.
            \end{itemize}
        \item disadvantages
            \begin{itemize}
                \item No proof of unbiasedness or consistency exists at the time of writing.
                \item $\mathcal{O}\bra{K^{2}}$ memory complexity.
                \item There is the need to choose a good kernel which may introduce extra parameters to train.
            \end{itemize}
    \end{itemize}
    
\section{Example usage}
\label{sec:ex_usage}
In this section we consolidate the example code presented throughout this package into a complete example demonstrating how to use \pkg{PyDPF} to run a simulated data DPF experiment. We show the full workflow: defining an SSM; simulating a dataset that obeys the given SSM; specifying a DPF algorithm; and finally training and testing the learned DPF. We also provide this example as \pkg{Jupyter} notebook tutorial on our GitHub\footnote{\url{https://github.com/John-JoB/pydpf/blob/main/tutorial_notebooks/pydpf-tutorial.ipynb}}.
\subsection{State space model}
In this example we use a toy, non-linear SSM for the returns of a financial asset taken from \citet{doucet2009tutorial}. Let $y_{0:T} \in \mathbb{R}^{T+1}$ be the realised returns of an asset and,
\begin{gather}
    \label{eq:ex_usage:stoch_vol_dyn}
    x_{t} = \alpha x_{t-1} + \sigma q_{t}, \forall t>0\\
    \label{eq:ex_usage:stoch_vol_obs}
    y_{t} = \beta \exp\bra{\frac{x_{t}}{2}}r_t,\\
    \label{eq:ex_usage:stoch_vol_pri}
    x_0 = \sqrt{\frac{\sigma^{2}}{1-\alpha^{2}}}, 
\end{gather}
where $x_{0:T}$ are the latent state, $q_{0:T},r_{0:T}$ are independent draws from a standard univariate Gaussian distribution, and $\alpha,\beta,\sigma$ are unknown parameters.

This model is a simplified discrete time stochastic volatility model. The latent state $x_{0:T}$ encodes the unobserved log volatility over a trajectory. $\beta^{2}$ is the volatility that the system tends to return to, known as the long-run volatility, as $\beta$ represents a standard deviation we restrict it to being positive. $\alpha$ encodes the system's tendency to return to the long-run volatility, therefore we expect $0<\alpha<1$. Finally, $\sigma^{2}$ is the volatility of log-volatility.

\subsection{Implementing the model}
For simplicity, we choose to use a bootstrap particle filter in this example. Therefore, neither the density of the \code{prior_model}, nor the \code{dynamic_model} are required to calculate the importance weights so we need not implement them, see Section \ref{sec:ssms:ssms}. We begin by implementing the \code{dynamic_model} with $\alpha$ and $\sigma$ registered as \pkg{PyTorch} \code{Module} parameters. We demonstrate the use of the \code{@pydpf.constrained_parameter} decorator through applying it to ensure that $0<\alpha<1$. We restrict $\sigma$ by parametrising it as $\log\sigma$ and efficiently recover $\sigma$ with a \code{@pydpf.cached_property}.

\footnotesize
\begin{Code}
class SVDynamicModel(pydpf.Module):

    def __init__(self, alpha, sigma, device, generator):
        super().__init__()
        self.alpha_ = torch.nn.Parameter(alpha)
        self.log_sigma = torch.nn.Parameter(torch.log(sigma))
        self.device = device
        self.generator = generator

    @pydpf.constrained_parameter
    def alpha(self):
        return self.alpha_, torch.clip(self.alpha_, 1e-3, 1-1e-3)

    @pydpf.cached_property
    def sigma(self):
        return torch.exp(self.log_sigma)

    def sample(self, prev_state, **data):
        state_size = prev_state.size()
        noise = torch.normal(0, 1, device=self.device, size=state_size, generator=self.generator)
        return prev_state * self.alpha + self.sigma * noise
\end{Code}
\normalsize
Next we define the \code{observation_model}. Filtering only accesses the \code{observation_model} to evaluate its log density for the importance weight calculation, however we will need to first simulate a dataset so we endow the \code{observation_model} with a \code{sample} method. We apply the same log parametrisation trick as with $\sigma$ in the \code{dynamic_model} to restrict $\beta$.

\footnotesize
\begin{Code}
class SVObservationModel(pydpf.Module):

    def __init__(self, beta, device, generator):
        super().__init__()
        self.log_beta = torch.nn.Parameter(torch.log(beta))
        self.half_log_2pi = torch.log(torch.tensor(2*torch.pi, device = device))/2
        self.device = device
        self.generator = generator

    @pydpf.cached_property
    def beta(self):
        return torch.exp(self.log_beta)

    def score(self, state, observation, **data):
        log_root_v = state + self.log_beta
        root_v = torch.exp(log_root_v)
        #Observations are independent of the particle so have one less
        #dimension than the particle dependent state, we unsqueeze
        #this dimension to broadcast over the particles.
        normalised_obs = observation.unsqueeze(1) / root_v
        return (-log_root_v - (normalised_obs**2)/2 - self.half_log_2pi).squeeze()

    def sample(self, state, **data):
        log_root_v = state + self.log_beta
        root_v = torch.exp(log_root_v)
        state_size = state.size()
        noise = torch.normal(0, 1, device=self.device, size=state_size, generator=self.generator)
        return root_v * noise
\end{Code}
\normalsize
Finally, we define the \code{prior_model}, as the \code{prior_model} and \code{dynamic_model} share parameters we pass the \code{dynamic_model} to the \code{prior_model}'s \code{__init__} function and make sure to not corrupt the computation graph by creating duplicate parameters.

\footnotesize
\begin{Code}
class SVPriorModel(pydpf.Module):

    def __init__(self, dynamic_model, generator):
        super().__init__()
        self.device = dynamic_model.device
        self.dyn_mod = dynamic_model
        self.generator = generator

    @pydpf.cached_property
    def sd(self):
        return torch.sqrt(self.dyn_mod.sigma**2 / (1-self.dyn_mod.alpha**2))

    def sample(self, batch_size, n_particles, **data):
        state_size = (batch_size, n_particles, 1)
        noise = torch.normal(0, 1, device=self.device, size=state_size, generator=self.generator)
        return self.sd * noise
\end{Code}
\normalsize
Now that we have defined all model components we require, we can group them together into a neat SSM object.
\footnotesize
\begin{Code}
def make_SSM(alpha, beta, sigma, device):
    dynamic = SVDynamicModel(alpha, sigma, device, generator)
    observation = SVObservationModel(beta, device, generator)
    prior = SVPriorModel(dynamic, generator)
    return pydpf.FilteringModel(prior_model=prior,
                                dynamic_model=dynamic,
                                observation_model=observation)
\end{Code}
\normalsize
\subsection{Generating synthetic data}
From a given state space model, simulating a large amount of trajectories is simple in \pkg{PyDPF}. We choose to simulate $200$ trajectories of $100$ time-steps from the model, \cref{eq:ex_usage:stoch_vol_dyn,eq:ex_usage:stoch_vol_obs,eq:ex_usage:stoch_vol_pri}, with $\alpha = 0.91, \beta = 0.5,\sigma = 1$. We simulate the trajectories in, parallelised if using \pkg{CUDA}, batches of 100.
\footnotesize
\begin{Code}
SSM = make_SSM(torch.tensor(0.91, device = device),
               torch.tensor(0.5, device = device),
               torch.tensor(1., device = device),
               device,
               generator)
#data_path must have a .csv extension
pydpf.simulate_and_save(data_path,
                        SSM = SSM,
                        time_extent = 100,
                        n_trajectories = 200,
                        batch_size = 100,
                        device = device)
\end{Code}
\normalsize
\subsection{Defining a DPF}
Before we can learn a filter, we have to specify the filter's functional form. In this example we use the soft resampling filter of \cite{karkus2018DPF}. We assume that the true parameters $\alpha,\beta,\sigma$ are unknown and initialise the algorithm with the guesses $\alpha = 0.6, \beta = 0.2, \sigma = 1.5$. For demonstration purposes, we show how to define a soft resampling DPF by specifying the resampling algorithm of an SIR particle filter, however in practice it is easiest to use the one-line method described in Section \ref{sec:imp:soft}.
\footnotesize
\begin{Code}
learned_SSM = make_SSM(torch.tensor(0.6, device = device),
                       torch.tensor(0.2, device = device),
                       torch.tensor(1.5, device = device),
                       device,
                       generator)
multinomial_base = pydpf.MultinomialResampler(generator = generator)
soft_resampler = pydpf.SoftResampler(softness = 0.7,
                                     base_resampler = multinomial_base,
                                     device = device)
DPF = pydpf.ParticleFilter(soft_resampler, learned_SSM)
\end{Code}
\normalsize
\subsection{Loading the data}
\label{sec:ex_usage_load}
We have described the \pkg{PyDPF} data loading process more completely in section \ref{sec:basic:data}. In this case the only data categories in our dataset are \code{state} and \code{observation}.
\footnotesize
\begin{Code}
full_dataset = pydpf.StateSpaceDataset(data_path,
                                       series_id_column = "series_id",
                                       state_prefix = "state",
                                       observation_prefix = "observation",
                                       device = device)
\end{Code}
\normalsize
Since \code{pydpf.StateSpaceDataset} subclasses \code{torch.Dataset} we may use the base \pkg{PyTorch} utility \code{torch.utils.data.random_split} to create train and test datasets.
\footnotesize
\begin{Code}
train_set, test_set = torch.utils.data.random_split(full_dataset, 
                                                    [0.5, 0.5], 
                                                    generator=cpu_gen)
\end{Code}
\normalsize
\pkg{PyDPF} uses the base \pkg{PyTorch} data loading functionality. The only special consideration is that, when using \code{torch.utils.data.random_split}, the \code{torch.utils.data.DataLoader} argument \code{collate_fn} must be set to the \code{collate} member of the \code{pydpf.StateSpaceDataset} that the passed dataset is a \code{torch.utils.data.Subset} of. Otherwise the data will not be returned in the expected order when the data loader is iterated over.
\footnotesize
\begin{Code}
train_loader = torch.utils.data.DataLoader(train_set, 
                                           batch_size=32, 
                                           shuffle=True, 
                                           collate_fn=full_dataset.collate,
                                           generator=cpu_gen)
test_loader = torch.utils.data.DataLoader(test_set, 
                                          batch_size=64, 
                                          shuffle=False, 
                                          collate_fn=full_dataset.collate)
\end{Code}
\normalsize
\subsection{Training a DPF}
\label{sec:ex_usage_train}
The \pkg{PyDPF} training procedure is very similar to base \pkg{PyTorch}. As in \pkg{PyTorch} we do not implement training loops in \pkg{PyDPF}. The exact form of the training loop will vary considerably depending on the experiment so we leave implementation up to the user. Moreover, we wish to allow a user to use \pkg{PyDPF} within their existing machine learning pipelines with minimal friction. The only difference between \pkg{PyDPF} and base \pkg{PyTorch} is that in \pkg{PyDPF} one should call the \code{.update()} method of all top-level \pkg{pydpf.Module} objects in addition to \code{optimizer.zero_grad()} between any optimiser step or backwards pass and the next forward pass. This ensures that the cache is properly invalidated so that cached values are correct and the computational graph is properly built.
\footnotesize
\begin{Code}
output_function = pydpf.MSE_Loss()
opt = torch.optim.Adam(DPF.parameters(), lr = 0.01)
n_epochs = 50
for e in range(n_epochs):
    train_loss = 0.0
    for state, observation in train_loader:
        opt.zero_grad()
        DPF.update()
        MSE = DPF(n_particles= 64,
                  time_extent=100,
                  aggregation_function=output_function,
                  observation=observation,
                  ground_truth=state)
        loss = MSE.mean()
        loss.backward()
        train_loss += loss.item()
        opt.step()
    if e 
        print(f"Epoch {e+1}, loss: {train_loss/len(train_loader)}")
    
    
DPF.update()
with torch.inference_mode():
    mean_loss = 0.0
    for state, observation in test_loader:
        MSE = DPF(n_particles= 64,
                      time_extent=100,
                      aggregation_function=output_function,
                      observation=observation,
                      ground_truth=state)
        mean_loss += MSE.mean().item()

print(f"Test MSE: {mean_loss/len(test_loader)}")
print(f"Learned alpha: {learned_SSM.dynamic_model.alpha.item()}")
print(f"Learned beta: {learned_SSM.observation_model.beta.item()}")
print(f"Learned sigma: {learned_SSM.dynamic_model.sigma.item()}")   
\end{Code}
\normalsize
For the reasons discussed in Section \ref{sec:basics:reproducibility}, we are unable to guarantee exact reproducibility across different hardware and software versions, however we provide the output of the above code block on our set-up as an example:
\footnotesize
\begin{Code}
Epoch 1, loss: 1.380717396736145
Epoch 11, loss: 0.8764162808656693
Epoch 21, loss: 0.7073744535446167
Epoch 31, loss: 0.6537422835826874
Epoch 41, loss: 0.6152108758687973

Test MSE: 0.6176820695400238
Learned alpha: 0.8957597017288208
Learned beta: 0.5009519457817078
Learned sigma: 1.0635138750076294
\end{Code}
\normalsize
\section{Advanced usage}
\label{sec:adv_usage}
\subsection{Conditional resampling}
We generally do not recommend using conditional resampling in differentiable filters. In a batch-parallel setting resampling is performed in parallel per-batch. The overhead required to evaluate the resampling criterion and partition the batch outweighs the cost benefit of not resampling all batches. 
However, conditional resampling is very commonly applied in classical particle filtering so we implement it in \pkg{PyDPF}.

When using \pkg{PyDPF} built-in filters, conditional resampling can be treated as just another resampling algorithm. One creates a \code{ConditionalResampler} module with a specified base resampler and condition, for example:

\footnotesize
\begin{Code}
    cond_resampler = pydpf.ConditionalResampler(
                     resampler=pydpf.MultinomialResampler(generator = gen),
                     condition=pydpf.ESS_Condition(threshold=0.7),
                     )
\end{Code}
\normalsize
The condition must be a callable object that takes the time-step data, (*state, *weight, etc.),  and returns a one dimensional boolean \code{Tensor} where true indicates a filter should be resampled, and false that it should not.

\subsection{Custom resamplers}
In many cases, including all those we have tutorialised in Section \ref{sec:imp-alg} aside from the marginal variant of the stop-gradient resampler, the only modifications made to the SIR particle filter, Algorithm \ref{alg:bg:sir_pf}, are to the resampling step, Step \ref{step:pf:resampling}. We therefore give special treatment to resampling and allow it to be modified apart from the rest of the particle filtering algorithm.

In \pkg{PyDPF} resamplers are \code{pydpf.Module} classes that implement a \code{.forward} method that takes the following data categories, see Table \ref{tab:data_cats}, \code{state}, \code{weight}, \code{observation}, \code{control}, \code{time}, \code{prev_time}, \code{series_metadata}, and \code{t}, as its input and returns a batch of resampled particle states and weights. For example one can implement the multinomial resampler as below:

\footnotesize
\begin{Code}
class MultinomialResampler(pydpf.Module):

    def __init__(self, generator:torch.Generator):
        super().__init__()
        self.generator = generator

    def forward(self, state, weight, **data):
        sampled_indices = torch.multinomial(torch.exp(weight), 
                                            weight.size(1), 
                                            replacement=True, 
                                            generator=self.generator).detach()
        return (pydpf.batched_select(state, sampled_indices), 
                    torch.full(weight.size(), -math.log(weight.size(1)), device=weight.device))
\end{Code}
\normalsize

\subsubsection{Returning additional information}
Occasionally, it may be required that the resampler returns additional information to the particle filter or \code{aggregation_function} either for diagnostics, as part of the loss function, or for use in an exotic filtering algorithm. What this information may be will depend on the specific filtering algorithm, for example one might want to track the particle genealogy for standard resampling algorithms, however genealogy is ill-defined for the optimal transport resampler. Furthermore, we cannot anticipate what information a user requires from custom resamplers. For this reason resamplers are permitted only to return the resampled \code{state} and \code{weight}. Any additional information should be stored in a \proglang{Python} dictionary at the \code{.cache} attribute.

A number of possible entries of \code{.cache} are used in the resamplers packaged with \pkg{PyDPF}. \code{.cache[`mask']} is defined for conditional resamplers and is a $1\text{D}$ boolean tensor of length $B$ where \code{True} indicates that a batch is resampled and \code{False} that it was not.

\code{.cache[`sampled_indices']} are the indices that the new particles are resampled from, \emph{e.g.} corresponding to $a^{(k)}_{t}$ in Algorithm \ref{alg:multi-res} for multinomial resampling. We register \\\code{.cache[`sampled_indices']} for all resamplers in \pkg{PyDPF} apart from \\ \code{pydpf.OptimalTransportResampler}.

\code{.cache[`used_weight']} are the weights of the atomic distribution the particles are simulated from, that may be different from the input particle weights, for example in soft-resampling. \code{cache[`used_weight']} should always be registered in any resampler.

The \pkg{PyDPF} implementation of the standard particle filter does not access\\ \code{.cache[`sampled_indices']} or \code{.cache[`used_weight']}, but many of the more complicated filters, \emph{e.g.} the marginal particle filter, do, so the user should register them in custom resampling algorithms if possible. We present the multinomial resampler including registering the \code{.cache} variables below:

\footnotesize
\begin{Code}
class MultinomialResampler(pydpf.Module):

    def __init__(self, generator:torch.Generator):
        super().__init__()
        self.generator = generator

    def forward(self, state, weight, **data):
        sampled_indices = torch.multinomial(torch.exp(weight), 
                                            weight.size(1), 
                                            replacement=True, 
                                            generator=self.generator).detach()
        self.cache['used_weight'] = weight
        self.cache['sampled_indices'] = sampled_indices
        return (pydpf.batched_select(state, sampled_indices), 
                 torch.full(weight.size(), -math.log(weight.size(1)), device=weight.device))
\end{Code}
\normalsize

\subsection{Custom filtering algorithms}
Some filtering algorithms are not examples of the SIR-PF, Algorithm \ref{alg:bg:sir_pf}, such as the interacting multiple model particle filter \citep{Boers2003IMM, brady2025DIMMPF}, and the marginal particle filter \citep{Klaas2005marginal, elvira2019auxiliaryMIS}. At the lowest level particle filters in \pkg{PyDPF} are implemented as iterated importance sampling. We allow the user to interact directly with this low level API. A new filtering algorithm can be defined by initialising \code{pydpf.SIS} with a callable object, \code{initial_proposal}, that takes the arguments [\code{n_particles}, \code{observation}, \code{t}, \code{control}, \code{series_metadata}, \code{time}] and returns the tuple (\code{state}, \code{weight}, \code{log-likelihood}); and a callable object, \code{proposal} that takes the arguments [\code{prev_state}, \code{prev_weight}, \code{observation}, \code{t}, \code{control}, \code{series_metadata}, \code{time}, \code{prev_time}] and returns the tuple (\code{state}, \code{weight}, \code{log-likelihood}). \code{log-likelihood} is the estimated log-likelihood factor $p\bra{y_{t}\mid y_{0:t-1}}$. 

We demonstrate this API with an example implementation of a bootstrap sequential importance sampler, \emph{i.e.} Algorithm \ref{alg:bg:sir_pf} without steps $6$-$10$.

\footnotesize
\begin{Code}
class BootstrapSISInitialProp(pydpf.Module):
    
    def __init__(self, prior_model, observation_model):
        super().__init__()
        self.prior_model = prior_model
        self.observation_model = observation_model
        
    def forward(self, n_particles, observation, **data):
        state = self.prior_model.sample(n_particles=n_particles,
                                        batch_size=observation.size(0), 
                                        **data)
        weights = self.observation_model.score(state=state, 
                                               observation=observation, 
                                               **data)
        normalised_weights, norm = pydpf.normalise(weights, dim=-1)
        return state, normalised_weights, norm - math.log(state.size(1))
    
class BootstrapSISProp(pydpf.Module):
    
    def __init__(self, dynamic_model, observation_model):
        super().__init__()
        self.dynamic_model = dynamic_model
        self.observation_model = observation_model
        
    def forward(self, prev_state, prev_weight, observation, **data):
        state = self.dynamic_model.sample(prev_state = prev_state, **data)
        score = self.observation_model.score(state=state, 
                                              observation=observation, 
                                              **data)
        normalised_weight, norm = pydpf.normalise(score + prev_weight, dim=-1)
        return state, normalised_weight, norm  - math.log(state.size(1))

custom_initial_proposal = BootstrapSISInitialProp(prior_model=custom_prior_model, 
                                                  observation_model=custom_observation_model)
custom_proposal = BootstrapSISProp(dynamic_model=custom_dynamic_model, 
                                   observation_model=custom_observation_model)
    
custom_filter = pydpf.SIS(initial_proposal=custom_initial_proposal, proposal=custom_proposal)
\end{Code}
\normalsize
For demonstration purposes we have shown the explicit passing of the custom proposal models to \code{pydpf.SIS}, but in practice it is neater to build a custom filter as a class that extends \code{pydpf.SIS} but overrides \code{.__init__()} to pass the custom implementations to \code{super().__init__()}.

\section{Package validation}
\label{sec:ex}
In this section we provide five examples that demonstrate the functionality of \pkg{PyDPF} and provide a comparison of the built-in DPFs. We first demonstrate our package's ability to perform vanilla (non-differentiable) particle filtering with a comparison to the Kalman filter. We contrast the time taken between our implementations when run solely on the CPU to when run with \pkg{CUDA} enabled GPU acceleration. Next, we test our package on a low-dimensional non-linear SSM, specifically a simplified stochastic volatility model \citep{doucet2009tutorial}. We use this example to demonstrate the complete \pkg{PyDPF} workflow, including defining a custom SSM, simulating data, loading data into a \code{dataset}, and training a DPF. We further use this example to investigate the variance of the gradient estimators, computational burden, and learning performance of the built-in algorithms. We then demonstrate an example usage of \pkg{PyDPF} to solve more complex, deep learning problems with a visual localisation example \citep{jonschkowski2018DPF}. Finally, we test the built-in algorithms on the challenging task of learning an efficient proposal distribution that has parameters not present in the parameter set of the SSM.

For all results reported in this paper, the CPU used is an Intel-i9-14900KF processor with $24$ cores and $64$GB of available RAM, and the GPU experiments are run with an Nvidia GeForce RTX $4090$ GPU with $24$GB of VRAM.

\subsection{Comparison with the Kalman filter}
    \label{sec:ex:lgssm}
    In this section, we will compare the Kalman filter against various particle filters applied to the linear-Gaussian state-space model.

    Linear-Gaussian state-space models have the form
        \begin{equation}
            \label{eq:lgssm_ex}
            \begin{aligned}
                \x_t &= \A \x_{t-1} + \q_t,\\
                \y_t &= \H \x_t + \r_t,
            \end{aligned}
        \end{equation}
    where $\A \in \mathbb{R}^{d_x \times d_x}$ is the state transition matrix, $\H \in \mathbb{R}^{d_y \times d_x}$ is the observation matrix, $\q_t \sim \mathcal{N}(\bm 0, \Q)$ is the state noise, and $\r_t \sim \mathcal{N}(\bm 0, \R)$ is the observation noise, and we have $\x_0 \sim \mathcal{N}(\bar{\x}_0, \mathbf{P}_0)$. In this model, the filtering distributions $p(\x_t|\y_{0:t})$ can be exactly recovered using the Kalman filter \citep{kalman1960new}.
    
    We present a modified variant of the linear-Gaussian set-up from \cite{naesseth2018variational, Corenflos2021OT}, altered in order to make the dynamic process stable, with $\mathbf{A}_{ij} = 0.38^{\mid i -j \mid +1}$, where $\mathbf{H} = \mathbf{I}_{d_y,d_x}$ is the matrix with ones on the diagonal for the first $d_y$ rows and zeros elsewhere, $\mathbf{Q}=\mathbf{I}_{d_x}$, $\mathbf{R}=\mathbf{I}_{d_y}$, $\bar{\x}_0 = \mathbf{0}_{dx}$, $\mathbf{P}_{0}= \mathbf{I}_{d_x}$. We assume that $d_y \leq d_x$.

    We compare the basic, non-differentiable particle filter implemented in PyDPF to the Kalman filter in Table \ref{tab:Kalman-Comparison}. We run both algorithms on $20$ batches of $100$ independently sampled trajectories with $T=1000$ time steps. We report the mean time per batch, in seconds; the mean squared error between the particle filter mean estimate and the exact Kalman mean; and the mean fractional error in the log-likelihood factors between the Kalman and particle filter estimates. Specifically the error in state estimate is defined as:
    \begin{equation}
    \label{eq:exp:epsilon_x}
        \epsilon_{x} = \frac{1}{TN}\sum^{T}_{t=0}\sum^{N}_{n=1}\mid\mid\hat{\x}^{(n,t)}_{\text{PF}}-\hat{\x}^{(n,t)}_{\text{Kalman}}\mid\mid^{2}_{2}\,,
    \end{equation}
    where $N=2000$ is the total number of trajectories, $\x^{i,j}_{\text{PF}}, \x^{i,j}_{\text{Kalman}}$ are the estimates of the latent state of the particle filter and Kalman filter respectively for trajectory $i$ and time-step $j$; and the fractional log-likelihood factor errors are defined as:
    \begin{equation}
    \label{eq:exp:epsilon_ell}
        \epsilon_{\ell} = \frac{1}{TN}\sum^{T}_{t=0}\sum^{N}_{n=1}\frac{\mid\hat{\ell}^{(n,t)}_{\text{Kalman}} - \hat{\ell}^{(n,t)}_{\text{PF}}\mid }{\hat{\ell}^{(n,t)}_{\text{Kalman}}}\,
    \end{equation}
    where $\hat{\ell}^{n,t}$ are the estimates of $p\bra{y_{t}\mid y_{0:t-1}}$.

    We compare run times for both the CPU and GPU. Refer to Section \ref{sec:ssms:pf} for instructions on running a particle filter, and Section \ref{sec:ssms:ssms} for instructions on defining a model in \pkg{PyDPF}.
    
    The idealised parallel time-complexity for the filtering algorithm on the GPU is $\mathcal{O}\bra{T\log K}$.
    We conjecture that the time differences across particle counts are largely due to memory management overheads; with higher particle counts the GPU will have to clear and reallocate memory more frequently.
    Additionally, the GPU has a finite number of threads available.
    The details of low level memory management are obscured by \pkg{PyTorch}, and therefore we cannot give a precise reason why the fastest time on GPU is achieved with $K=10^{3}$ particles rather than the expected $K=25$ particles as on the CPU. We believe this is due to \pkg{CUDA} more efficiently chunking larger tensors.

    \begin{table}[th]
        \centering
        \begin{tabular}{|c|c|c|c|c|}
            \hline
              & Time CPU $(s)$ & Time GPU $(s)$& $\epsilon_{x}$ & $\epsilon_{\ell}$ \\
             \hline
             Kalman Filter & $0.64$ & $2.3$ & $0$ & $0$ \\
             PF $K=25$ & $1.1$ & $1.3$ & $3.8$ & $0.14$ \\
             PF $K=10^{2}$ & $3.0$ & $1.5$ & $1.1$ & $0.071$ \\
             PF $K=10^{3}$ & $34$ & $0.65$ & $0.11$ & $0.022$ \\
             PF $K=10^{4}$ & $360$ & $5.1$ & $0.012$ & $0.0071$ \\
             \hline
        \end{tabular}
        \caption{A comparison of a \pkg{PyDPF} particle filter against the Kalman filter for a linear Gaussian model with $d_x = 25, \, d_y = 1$.}
        \label{tab:Kalman-Comparison}
    \end{table}

\subsection{Performing filtering given a fully specified model: stochastic volatility}
\label{sec:ex:filtering_only}
We use the same stochastic volatility example, \cref{eq:ex_usage:stoch_vol_dyn,eq:ex_usage:stoch_vol_obs,eq:ex_usage:stoch_vol_pri}, as we used to tutorialise the package in \ref{sec:ex_usage}. In this section we demonstrate the ability of our package to filter non-linear systems with a fully specified SSM.

We compare the performance of the implemented filters in Table \ref{tab:fully_spec} where $\epsilon_x$ and $\epsilon_\ell$ are as in Eqs. \eqref{eq:exp:epsilon_x} and \eqref{eq:exp:epsilon_ell} respectively, except instead of comparing with the Kalman filter we have compared to an SIR particle filter (Alg. \ref{alg:bg:sir_pf}) with $10^{4}$ particles. $\epsilon_x$ and $\epsilon_\ell$ for the non-differentiable, stop-gradient, and marginal stop-gradient filters are identical. This is because, for bootstrap filtering, these algorithms are equivalent on the forward pass. These three algorithms also report near identical times as \pkg{PyDPF} will automatically recognise that gradient is not tracked so short circuits any unneeded computation. Soft gradient resampling with $\xi = 0.7$ makes little difference to the accuracy in this example. Optimal transport resampling, with $\epsilon=0.5$, and kernel resampling, with a Gaussian kernel of bandwidth $0.1$, lose some accuracy compared to the other methods. The forward pass time is similar for all filters, except the optimal transport filter which is much slower.

\begin{table}[th]
    \footnotesize
        \centering
        \begin{tabular}{|c|c|c|c|}
            \hline
              Resampling method  & $\epsilon_{x}$ & $\epsilon_{\ell}$ & Forward Time $(s)$\\
             \hline
             Non-differentiable &  0.027 & 0.064 & 0.62\\
             Soft & 0.026 & 0.065 & 0.74\\
             Stop-gradient & 0.027 & 0.064 & 0.64 \\
             Marginal stop-gradient & 0.027 & 0.064 & 0.64 \\
             Optimal transport & 0.049 & 0.079 & 26 \\
             Kernel-mixture & 0.033 & 0.073 & 0.62\\
             \hline
        \end{tabular}
        \caption{A comparison of DPFs included in \pkg{PyDPF} on a simple non-linear stochastic volatility model, in the time taken to complete the forward passes for a batch of $128$ trajectories, as well as the $\epsilon_x$ and $\epsilon_\ell$ using a particle filter with $K=10^{4}$ as the reference. $T=1000$. $K=100$.}
        \label{tab:fully_spec}
    \end{table}

\subsection{Unsupervised learning of a single parameter: stochastic volatility}
    \label{sec:ex:learn_one_param_small}
    In Section \ref{sec:ex:filtering_only} we performed filtering assuming that all parameters of the model given in \cref{eq:ex_usage:stoch_vol_dyn,eq:ex_usage:stoch_vol_obs,eq:ex_usage:stoch_vol_pri} are known.
    However, if this is not the case, then we must estimate the unknown parameters before we can perform filtering. In this example, we assume that only the $\alpha$ parameter of Eq.~\eqref{eq:ex_usage:stoch_vol_dyn} is unknown, and demonstrate various methods for estimating this parameter.
    
    We note that low dimensional parameter estimation in state-space models is not a primary usage of \pkg{PyDPF}; the classical methods discussed in \cite{kantas2015parameter-est} are typically more suitable, but we include this simple example to illustrate a basic workflow. Example code to load and train a DPF on a dataset in \pkg{PyDPF} can be found in Section \ref{sec:ex_usage_load} and \ref{sec:ex_usage_train} respectively.

    This example is too simple to meaningfully discriminate between the algorithms by the performance of the learned model or by the distance to the true $\alpha$. We use this simple example to compare the time taken and the standard deviation of the gradient estimate for a single random observation trajectory from \cref{eq:ex_usage:stoch_vol_dyn,eq:ex_usage:stoch_vol_obs,eq:ex_usage:stoch_vol_pri}. In the interest of reproducibility we have uploaded the specific trajectory used to our GitHub repository as \code{/jss_examples/Stochastic volatility/test_trajectory.csv}.

    We choose to generate data from the model with $\alpha = 0.91,\,\beta = 0.5,\, \sigma=1$. In our experiments soft resampling has $\xi=0.7$, optimal transport resampling has $\epsilon=0.5$; and kernel-mixture resampling uses Gaussian kernels with a variance of $0.3$.

    We present the results of this experiment in Table \ref{tab:Single-parameter}. As in Table \ref{tab:Kalman-Comparison}, we observe unintuitive timings where more operation-expensive algorithms are slightly faster to run than theoretically cheaper ones. 

    \begin{table}[th]
    \footnotesize
        \centering
        \begin{tabular}{|c|c|c|c|c|}
            \hline
              Resampling method & Forward Time $(s)$& Backward Time $(s)$ & Gradient s.d. & $\alpha$ abs. error\\
             \hline
             Non-differentiable &  0.19 & 0.089 & 0.035 & 0.0039 \\
             Soft & 0.22 & 0.13 & 0.38 & 0.0062\\
             Stop-gradient & 0.20 & 0.11 & 1.2 & 0.0056\\
             Marginal stop-gradient & 0.13 & 0.058 & 0.48 & 0.0047 \\
             Optimal transport & 2.7 & 0.099 & 13 & 0.073 \\
             Kernel-mixture & 0.097 & 0.0062 & 0.35 & 0.0020\\
             \hline
        \end{tabular}
        \caption{A comparison of DPFs included in \pkg{PyDPF} in the time taken to complete the forward and backward passes for a batch of $100$ trajectories; the standard deviation of the gradient of the ELBO divided by the number of timesteps of a single trajectory over $2000$ repeats with $\alpha$ frozen at $0.93$; and the mean absolute error in the learned parameter $\alpha$ after 10 training epochs with $500$ total trajectories  $T = 100$. $K=100$.}
        \label{tab:Single-parameter}
    \end{table}

\subsection{Unsupervised learning of multiple parameters: stochastic volatility}
    \label{sec:ex:learn_multi_param_small}
    In Section \ref{sec:ex:learn_one_param_small}, we demonstrated how to utilise our package to infer the value of the $\alpha$ parameter of \cref{eq:ex_usage:stoch_vol_dyn,eq:ex_usage:stoch_vol_obs,eq:ex_usage:stoch_vol_pri} given that the $\beta$ and $\sigma$ parameters are known.
    
    In this example, we will assume that $\alpha, \beta,$ and $\sigma$ are unknown in \cref{eq:ex_usage:stoch_vol_dyn,eq:ex_usage:stoch_vol_obs,eq:ex_usage:stoch_vol_pri}, we will learn them by optimising the ELBO for each of our implemented DPFs. We present the results in Table \ref{tab:Multiple-parameter-simple}. 
    
    The main conclusion is that, for this simple model and utilising the bootstrap proposal, the low-variance attained by not accumulating gradient over time-steps is preferential to the less biased algorithms.

    \begin{table}[th]
    \footnotesize
        \centering
        \begin{tabular}{|c|c|c|c|c|}
            \hline
              Resampling method & Test ELBO & $\alpha$ abs. error & $\beta$ abs. error & $\sigma$ abs. error \\
             \hline
             Non-differentiable &  -106.4 & 0.0050 & 0.036 & 0.037\\
             Soft & -106.6 & 0.0092 & 0.17 & 0.067\\
             Stop-gradient & -106.3 & 0.0045 & 0.09 & 0.054 \\
             Marginal stop-gradient & -106.4 & 0.0057 & 0.14 & 0.051 \\
             Optimal transport & -108.5 & 0.075 & 0.74 & 0.33 \\
             Kernel-mixture & -106.6 & 0.014 & 0.24 & 0.088\\
             \hline
        \end{tabular}
        \caption{A comparison of DPFs included in \pkg{PyDPF} by their achieved test ELBO and L1 error in the learned parameters. We used $500$ total trajectories with a train:validation:test split of $2:1:1$ and a batch size of $30$. We took $T = 100$, $K=100$ during training and $K=1000$ during testing. All statistics are averaged over 10 independent datasets and training runs of $20$ epochs. The parameters were initialised from a uniform distribution in the ranges: $[0,1]$ for $\alpha$, $[0,2]$ for $\beta$, $[0,5]$ for $\sigma$. These parameters were optimised by stochastic gradient descent with a learning rate equal to $1/10$ of their initial range that was exponentially decayed by a factor of $0.95$ each epoch.}
    \label{tab:Multiple-parameter-simple}    
    \end{table}

\subsection{Deep learning: visual localisation}

    Visual localisation has been the application where DPFs have received the greatest research attention \cite{jonschkowski2018DPF, karkus2018DPF, younis2024diff-smoother}. We use trajectories and paired images simulated in DeepMind lab \citep{beattie2016deepmindlab}. This task was introduced to test DPFs by \cite{jonschkowski2018DPF} and has remained popular and has been used in \cite{chen2024normalisingflow}, \cite{Corenflos2021OT}, \cite{li2024semi-sup}, and \cite{younis2024mixtureresample} amongst other works. Specifically we follow the experimental set-up from \cite{chen2024normalisingflow} and use their neural network architectures. This example demonstrates the workflow for using \pkg{PyDPF} to address complicated deep learning problems and demonstrates the performance of our implemented algorithms.
    
    The goal is to, given a series of images from a forward facing camera of a simulated robot and a series of control actions, estimate the location of the robot at some late time-step. We approach this problem as a supervised learning task, where we assume we have access to the ground truth position at all time-steps for all trajectories. We assume no knowledge of the robot's starting position in the maze.

    The Maze has dimensions of $2000\times1300$ and we use trajectories of $100$ time-steps. The images from the front-facing camera are $32\times32\times3$ RGB, we randomly crop them to $24\times24$ with the top-left pixel of the retained section being uniformly chosen from $\bra{0,\dots,7; 0,\dots,7}$. The control actions are deterministic and exactly equal to the true change in state, but given in the frame of the robot.

    \subsubsection{Prior model}
    We initialise the particles randomly and uniformly over the area of the maze.

    \subsubsection{Observation model}
    The observations are encoded into a $128$ dimensional feature vector via a convolutional neural network of three convolutional layers and one linear layer for a total of $115,808$ trainable parameters. We similarly transform the particle locations to a $64$ dimensional feature vector through four-layer multilayer perceptron with a total of $6,896$ trainable parameters. We calculate a scoring value from the encoded state and observation through a simplified normalising flow model:
    \begin{subequations}
    \label{eq:ex:normflow}
    \begin{gather}
        F\bra{\mathbf{z}} := \text{concat}\bra{\bm \alpha, \bm \beta}\,,\\
       \bm \alpha := F_{\text{L}}\bra{\z_{\text{U}}; \x, \bm \theta} + \z_{\text{L}}\,,\\
      \bm \beta :=F_{\text{U}}\bra{\bm \alpha; \x, \bm \theta} + \z_{\text{U}}\,,
    \end{gather}
    \end{subequations}
    where $\z_{\text{L}}$ and $\z_{\text{U}}$ are the first and latter halves of the random vector $\z$. We assume that we can simulate and evaluate the density of $\z$ in a differentiable manner; $\text{concat}\bra{\cdot,\cdot}$ is the vector concatenation operation; $F_{\text{L}}$ and $F_{\text{U}}$ are arbitrary differentiable functions; and $\x$ is a conditioning variable. Notice that $F\bra{\cdot}$ is invertible and has a Jacobian determinant of $1$, so given $\x, \bm \theta$ we can both simulate from $\y \sim F\bra{\z}$ and evaluate the density of a given $\y$.

    We model the observations as:
        
    \begin{subequations}
    \label{eq:ex:maze_obs_model}
    \begin{gather}
        E_{\text{obs.}}\bra{\y_{t}; \bm \theta} = F_{1}\bra{F_{2}\bra{\z_{t}; E_{\text{state}}\bra{\x_{t}; \bm \theta}, \bm \theta}; \bm \theta}, \\
        z \sim \mathcal{N}\bra{0, 1},
    \end{gather}
    \end{subequations}
    where $F_{1}\bra{\cdot}$ and $F_{2}\bra{\cdot}$ are normalising flows, Eq. \eqref{eq:ex:normflow}, with different and independently parameterised multilayer perceptrons $F_{\text{L}}$, $F_{\text{U}}$; $E_{\text{obs.}}\bra{\cdot; \bm \theta}$ and $E_{\text{state}}\bra{\cdot; \bm \theta}$ are the observation and state encoders respectively. Together, $F_{1}$ and $F_{2}$ have a total of $181,504$ trainable parameters.

    Note that the probability returned by calculating the density under the model Eq. \eqref{eq:ex:maze_obs_model} is not a true likelihood on the observations as it does not account for the transformation through the non-bijective function $E_{\text{obs.}}\bra{\cdot; \bm \theta}$.

    \subsubsection{Dynamic model}
    The dynamic model transforms the state by the control action with additive Gaussian noise.
    \begin{gather}
        x_{1,t} = c_{1,t}\cos x_{3,t-1} - c_{2,t}\sin x_{3,t-1} + x_{1,t-1} +\epsilon_{1,t}\,, \\
        x_{2,t} = c_{1,t}\sin x_{3,t-1} + c_{2,t}\cos x_{3,t-1} + x_{2,t-1} +\epsilon_{2,t}\,, \\
        x_{3,t} = x_{3,t-1} + \epsilon_{3,t}\,,\\
        \bm \epsilon \sim \mathcal{N}\bra{\mathbf{0}, \text{diag}\bra{\sqrt{30}, \sqrt{30}, \sqrt{0.3}}}\,.
    \end{gather}

    \subsubsection{Training loss}
    We use a very similar loss to \cite{chen2024normalisingflow}, despite only using the last time-step for validation we minimise the MSE across the entire trajectory during training.
    \begin{equation}
        \mathcal{L}_{\text{MSE}} = \frac{1}{T}\sum^{T}_{t=0}\norm{\frac{1}{1000}\bra{\sum^{K}_{k=1}\bar{w}^{(k)}_{t}\bra{x^{(k)}_{1,t},  x^{(k)}_{2,t}}^{T} - \bra{\bra{x_{\text{GT}}}_{1,t}, \bra{x_{\text{GT}}}_{2,t}}^{T}}}^{2}_{2}\,, 
    \end{equation}
    where $\bra{\x_{\text{GT}}}_{t}$ is the ground truth state.

    We also penalise inaccuracy in the estimated orientation of the robot.
    
    \begin{equation}
    \begin{gathered}
        \ell_t(\overline{w}, \x, \x_{\text{GT}}) = \sum^{K}_{k=1}\bar{w}^{(k)}_{t}\bra{\sin \bra{x^{(k)}_{3,t}},  \cos \bra{x^{(k)}_{3,t}}}^{T} - \bra{\sin\bra{\bra{ x_{\text{GT}}}_{3,t}}, \cos\bra{\bra{x_{\text{GT}}}_{3,t}}}^{T}\\
        \mathcal{L}_{\text{Angle}} = \frac{1}{T} \sum_{t=0}^{T} \norm{\ell_t(\overline{w}^{(1:K)}_{t}, \x^{(1:K)}_{t}, \bra{\x_{\text{GT}}}_{t})}^{2}_{2}
    \end{gathered}
    \end{equation}
    Following the recommendation in \cite{li2024semi-sup}, to help the encoder learn the features of the observation we define a decoder and employ an auto-encoder loss:
    \begin{equation}
        \mathcal{L}_{\text{AE}} = \frac{1}{Td_{\y}}\sum^{T}_{t=0}\norm{\y_{t} - D_{\text{obs.}}\bra{E_{\text{obs.}}\bra{\y_{t}; \bm \theta}; \bm \theta}}^{2}_{2}\,,
    \end{equation}
    where $D_{\text{obs.}}\bra{\cdot;\bm \theta}$ is the observation decoder and $d_{\y}$ is the dimension of the observations.

    The training objective is written
    \begin{equation}
        \mathcal{L}_{\text{Training}} = \mathcal{L}_{\text{MSE}} + \mathcal{L}_{\text{Angle}} + \mathcal{L}_{\text{AE}}\,.
    \end{equation}

    The results for this experiment are given in Table \ref{tab:deep-mind}. To evaluate the filters we compare the MSE at only the last time-step, this is because the observations and the prior are only weakly informative so to accurately localise the agent the algorithm needs information from several images along the trajectory. We report the time elapsed over the complete training-validation-testing procedure as this represents the practical cost to deploy each algorithm. Due to the use of \pkg{CUDA} convolutional layers, this experiment returns significantly different results across runs even if the random seed is held constant under the default environment settings. We run the experiment under the \code{pydpf.utils.set_deterministic_mode(True, True)} context to limit non-deterministic behaviour. Because the deterministic mode induces a slowdown we additionally time the algorithms with the default settings.

    The best performance was seen using the soft-resampler, $\xi = 0.7$, mirroring the results from \cite{chen2024normalisingflow}. We conjecture that it outperforms the stop-gradient and marginal stop-gradient methods due to its lower variance despite the additional bias. The non-differentiable resampling also has low variance but unlike soft-resampling it does not pass any gradient information through time-steps so struggles to capture the dependencies between time-steps.

    As in \cite{chen2024normalisingflow}, we were unable to find a hyper-parameter setting for optimal-transport resampling that was stable enough to converge. Moreover, its training and inference costs are very high in comparison to the other techniques. In \cite{Corenflos2021OT} this resampler is successfully applied to the DeepMind maze environment but under a considerably less challenging set-up.

    We find that kernel-mixture resampling performs poorly, however our training and evaluation targets differ from those used in the paper that proposed it \citep{younis2024mixtureresample}.

    \begin{table}[th]
    \footnotesize
        \centering
        \begin{tabular}{|c|c|c|c|c|}
            \hline
              Resampling method & $\sqrt{\text{Test MSE}}$ & Deterministic time (hrs:mins) & Non-deterministic time (hrs:mins) \\
             \hline
             Non-differentiable & 302 & 00:13 & 00:12\\
             Soft & 247 & 00:15 & 00:13\\
             Stop-gradient & 377 & 00:14 & 00:12 \\
             Marginal stop-gradient & 357 & 00:14 & 00:14\\
             Optimal transport & 1212 & 01:32 & 01:37\\
             Kernel-mixture & 565 & 00:15 & 00:14\\
             \hline
        \end{tabular}
        \caption{A comparison of DPFs included in \pkg{PyDPF} by the square root of their achieved  test MSE at the last time-step and the total time to complete a training-validation-testing run for the deep mind maze set-up. We use $900$ training trajectories, $400$ for validation, and $700$ for testing with a batch size of $64$ over $100$ training epochs. $T=99$, $K=100$. All results are reported as an average over $5$ independent training runs. The square root MSEs reported are from running the model in deterministic mode.}
        \label{tab:deep-mind}
    \end{table}

\subsection{Learning proposal parameters}
\label{sec:ex:prop}
In this section we demonstrate the ability of the algorithms implemented in \pkg{PyDPF} to learn efficient proposal distributions. We use the same simple SSM as in Section \ref{sec:ex:lgssm}. We parameterise the proposal model with
\begin{equation}
\label{eq:ex:prop:prop}
    x_{t} \sim q_{\text{Learned}} = \mathcal{N}\bra{\mathbf{G}\mathbf{A}\x_{t-1} + \mathbf{H}\y_{t}, \mathbf{S}}\,,
\end{equation}
where $\mathbf{G}, \mathbf{S}$ are $d_x \times d_x$ matrices with the leading diagonals being the learned parameter vectors $\mathbf{\phi}_{\mathbf{G}}$ and $\mathbf{\phi}_{\mathbf{S}}$, respectively and all other elements being set to zero; and $\mathbf{H}$ is a $d_x \times d_y$ matrix with the leading diagonal being the learned parameter vector $\mathbf{\phi}_{\mathbf{H}}$ and all other elements being set to zero.

The locally optimal proposal, $q_{\text{Opt}}$, that minimises the variance of the weights at time-step $t$ given a resampled particle at time-step $t-1$ \citep{chopin2020book}, is included in the parameterised family given by Eq. \eqref{eq:ex:prop:prop}, with \begin{gather}
    \bra{\phi_{\mathbf{G}}}_{i} = \bra{\phi_{\mathbf{S}}}_{i} = 
    \begin{cases}
        \frac{1}{2} \;\; i \leq d_{y},\\
        1 \;\; i > d_y,
    \end{cases}\\ 
    \bra{\phi_{\mathbf{H}}}_{i} = \frac{1}{2}\,.
\end{gather}
We optimise the SMC ELBO, simultaneously developed in \cite{naesseth2018variational, le2018auto-encoding, maddison2017variational}. The locally optimal proposal is not guaranteed to coincide with the optimal ELBO, however it has been found experimentally that 
optimising the proposal with respect to the ELBO improves sampling efficiency \citep{Cox2024GaussMix, Corenflos2021OT}.
We refer the reader to \cite{naesseth2018variational, le2018auto-encoding} for theoretical discussion.

We evaluate the quality of the learned proposals on four metrics. Firstly, the accuracy of the learned filter, $\epsilon_{x}$ and $\epsilon_{\ell}$ defined in Eqs. \eqref{eq:exp:epsilon_x} and \eqref{eq:exp:epsilon_ell}.
In order to evaluate how close the learned proposal is to the locally optimal proposal, for each repeat we calculate the squared 2-Wasserstein distance between the optimal proposal and the learned proposal averaged over every time-step in the test dataset. We report the square-root average over the repeats of each averaged distance:
\begin{equation}
    \begin{gathered}
    D_{\mathcal{W}} = \sqrt{\frac{1}{R}\sum^{R}_{r=1}\sbra{\frac{1}{(T+1)S}\sum_{\x_{t-1},\y \in \text{test dataset}}\inf_{Q^{(r)}}\mathbb{E}_{(\x_{t}, \x'_{t}) \sim Q^{(r)}}\sbra{\lVert \x_{t} - \x'_{t}\rVert^{2}_{2}}}}\,, \;\;  \\\text{s.t.} \;\; Q^{(r)}\bra{\x_{t},\x'_{t}} \: \text{has the marginals} \: \x_{t}\sim \bra{q_{\text{Learned}}}^{(r)}, \x'_{t}\sim q_{\text{Opt}},
\end{gathered}
\end{equation}
where $S$ is the number of independent trajectories in the test dataset and $\bra{q_{\text{Learned}}}^{r}$ is the learned distribution at experiment repeat $r$ with a given $\x_{t-1}$ and $\y_{t}$. 
\begin{table}[th]
    \small
        \centering
        \begin{tabular}{|c|c|c|c|c|}
            \hline
              Resampling method & $\epsilon_{x}$ & $\epsilon_{\ell}$ & $D_{\mathcal{W}}$ & ELBO \\
             \hline
             Bootstrap & 0.89 & 0.068 & 0.77 & -1803.2\\
             Locally Optimal & 0.39 & 0.018 & 0.0 & -1797.7 \\
             Non-differentiable & 0.44 & 0.030 & 0.22 & -1798.1\\
             Soft & 0.39 & 0.019 & 0.035 & -1797.7\\
             Stop-gradient & 0.46 & 0.035 & 0.27 & -1798.4 \\
             Marginal stop-gradient & 0.43 & 0.029 & 0.22 & -1798.0\\
             Optimal transport & 0.46 & 0.058 & 0.47 & -1800.4\\
             Kernel-mixture & 1.31 & 0.10 & 0.89 & -1805.9\\
             \hline
        \end{tabular}
        \caption{A comparison of DPFs included in \pkg{PyDPF} on their ability to learn an efficient proposal distribution. We use $T=1000$ and $K=100$ and take a batch size of $32$ during training. The results are averaged over $5$ repeats.}
        \label{tab:proposal-learning}
    \end{table}

\section{Conclusion}
\label{sec:conc}
    This paper has introduced \pkg{PyDPF}, a software package unifying several differentiable particle filters, allowing them to be used to perform inference in state-space models.
    We have described several of the implemented algorithms, and provided multiple examples of the usage of our package in practical examples of various difficulty.
    Our package is designed to be flexible, and to easily interoperate with \pkg{PyTorch}, allowing for efficient usage of modern deep learning methods within a state-space model context.
    Furthermore, our package leverages GPU computing via \pkg{PyTorch}, allowing for fast parallel evaluation of particle filters on non-interacting trajectories.
    Finally, our package is designed to be extensible, allowing for rapid design and implementation of novel particle filtering methods within the provided framework.

\section*{Acknowledgments}
    We thank Xiongjie Chen for his assistance in setting up the deep mind maze environment and Zheng Zhao for useful discussion.
    JJ. Brady acknowledges support from the National Physical Laboratory of the United Kingdom via an NMI/EPSRC studentship.
    J. Cox acknowledges support from Germany’s Federal Ministry of Research, Technology and Space (BMFTR) within the ErUM-Data programme under grant FKZ 05D25PC1 (DEMOS consortium). 

\bibliography{ref}
\end{document}